\documentclass{aa}  

\usepackage{graphicx}
\usepackage{txfonts}
\usepackage{hyperref}
\usepackage{natbib}
\usepackage{longtable}
\bibpunct{(}{)}{;}{a}{}{,} 
\usepackage[dvipsnames]{xcolor}

\newcommand{\micron}{\ensuremath{\mu}m}

\newcommand{\re}{\ensuremath{R_\textrm{e}}\xspace}

\newcommand{\kms}{km\,s\ensuremath{^{-1}}\xspace}%
\newcommand{\msun}{\ensuremath{M_{\sun}}\xspace}

\newcommand{\mbh}{\ensuremath{M_{\bullet}}\xspace}

\newcommand{\vlos}{\ensuremath{V_\textrm{LOS}}\xspace}

\newcommand{\slos}{\ensuremath{\sigma_{\textrm{LOS}}}\xspace}
\newcommand{\sgra}{Sgr~A$^\star$\xspace}

\newcommand{\qmin}{\ensuremath{q_\textrm{min}}\xspace}
\newcommand{\pmin}{\ensuremath{p_\textrm{min}}\xspace}
\newcommand{\umin}{\ensuremath{u_\textrm{min}}\xspace}
\newcommand{\ups}{\ensuremath{\Upsilon}\xspace}
\newcommand{\lz}{\ensuremath{\lambda_z}\xspace}
\begin{document}

\title{A spectroscopic map of the Galactic centre}

 \subtitle{Integrated light and dynamical modelling}

 \author{A. Feldmeier-Krause 
 \inst{1}
 \and
T.~I.  Maindl\inst{1}\fnmsep\inst{2}
 \and G. van de Ven\inst{1}
 \and S. Thater\inst{1}
 \and P. Jethwa\inst{1}
 \and I. Breda\inst{1}\fnmsep\inst{3}
 }

 \institute{Department of Astrophysics, University of Vienna, T\"urkenschanzstrasse 17, 1180 Wien, Austria\\
 \email{anja.krause@univie.ac.at}
         \and
             SDB Science-driven Business Ltd., Faneromenis Avenue 85, Ria Court 46, Suite 301, 6025 Larnaca, Cyprus
            \and
            Instituto de Astrofísica e Ciências do Espaço, Universidade de Lisboa, OAL, Tapada da Ajuda, 1349-018 Lisboa, Portugal
             }

   \date{Received March 13, 2026; accepted May 03, 2026}

  \abstract
   {The centre of the Milky Way is occupied by a dense nuclear star cluster that contains the supermassive black hole \sgra. The mildly flattened cluster is embedded in the larger surrounding nuclear stellar disc. These three components dominate the mass budget of the Galactic centre at different radial scales.  }
   {The mass distribution of the Galactic centre has been studied extensively using observations of individual bright stars and various dynamical modelling approaches. The situation differs for external galaxies, where observations are often limited to the integrated line-of-sight kinematics. For such systems, triaxial orbit-based dynamical modelling has become a standard method of deriving mass distributions and stellar orbit distributions. We aim to apply and test this method on the Galactic centre.}
   {We extracted stellar line-of-sight kinematic maps of the inner $\sim$3\,pc$\times$66\,pc region of the Galactic centre. We used the \textsc{dynamite} code, which calculates an orbit library in a given gravitational potential and computes model kinematic maps. These model maps were then compared to the observed kinematic maps, and thus the gravitational potential and orbit distribution of the Galactic centre were constrained.}
   {We recover the correct mass of \sgra, and our stellar mass distributions are in agreement with the literature, albeit with larger uncertainties. We find that the stellar structures are at most mildly triaxial and close to oblate. The contribution of dark matter to the total mass distribution is of the order of $\textless$1\%. The stellar orbit distribution in the inner $\sim$33\,pc region is dominated by dynamically warm and hot orbits. At larger scales of $\sim$80--160\,pc, dynamically cold --  highly rotating --  orbits have the largest weights.  }
   {The dominance of hot and warm orbits is a consequence of short dynamical timescales in the inner Galactic centre, causing dynamical heating of the orbits. The presence of cold orbits at large radii may be explained by the longer heating timescales in this region, and by the stars in the outer nuclear stellar disc being younger. 
The agreement of our mass distribution with other studies confirms the validity of the orbit-based modelling approach. } 

   \maketitle

\section{Introduction}

Galaxy centres play an important role in the formation and evolution of galaxies. Inflow and outflow of gas to and from the galaxy centre is common, and found in about 50\% of nearby low-ionisation nuclear emission-line regions \citep[LINERs,][] {2022A&A...660A.133H}. Although galaxy centres only extend over a small spatial scale of the galaxy (up to a few hundred parsecs), tight scaling relations between properties of the galaxy centre and the more extended galaxy have been found. These scaling relations imply that the large-scale galaxy assembly and the build-up of the galaxy centre are connected.  For example, there is a correlation between the mass of the supermassive black hole (\mbh) and the host galaxy stellar velocity dispersion \citep[e.g.][]{2000ApJ...539L...9F,2000ApJ...539L..13G,2013ARA&A..51..511K}, and a correlation between the mass of the nuclear star cluster  (NSC) and the galaxy stellar mass \citep[e.g.][]{2006ApJ...644L..21F,2013ApJ...763...76S,2016MNRAS.457.2122G}.  The slope of these relations, their scatter, and behaviour at different mass regimes and environments are still areas of active research.

The most reliable way to measure an NSC mass is via stellar dynamical modelling \citep{2020A&ARv..28....4N}, using stellar kinematic data. 
In most NSCs, we can so far only use the spectroscopic integrated stellar light for dynamical modelling \citep{2009ApJ...690.1031B,2018ApJ...858..118N}. The Milky Way (MW)'s centre is one of the few galaxy centres where we can resolve single stars and use those to constrain the mass distribution. 
This makes the Galactic centre (GC) an excellent benchmark object where we can test the methods and techniques used to infer the mass distributions in other galaxies. 

The GC contains several massive structures, each of which dominates the total gravitational potential at a certain scale: (1) the mildly flattened NSC with an effective radius, \re, of about 5\,pc \citep{2016ApJ...821...44F,2020A&A...634A..71G} and a total mass of a few $10^7$\,\msun \citep[e.g.][]{2014A&A...570A...2F,2015MNRAS.447..948C,2017MNRAS.466.4040F}; (2) the more flattened surrounding nuclear stellar disc (NSD) with a scale radius of about 100\,pc, a scale height of $\sim$30\,pc \citep{1999A&A...348..768P,2002A&A...384..112L,2022MNRAS.512.1857S}, and a mass of $\sim$$10^9$\,\msun \citep{2022MNRAS.512.1857S}; and (3) the central supermassive black hole \sgra. The mass of \sgra\ is known with high precision, \mbh=(4.30$\pm$ 0.012)$\cdot$10$^6$\,\msun, \citep{2022A&A...657L..12G}, thanks to the long-time monitoring of stellar orbits \citep[e.g.][]{2016ApJ...830...17B,2017ApJ...837...30G,2019Sci...365..664D}. 
The mass distributions of the MW's NSC and NSD have been studied with various dynamical modelling approaches, some of which assume spherical symmetry \citep[e.g.][]{
2019MNRAS.484.1166M}, axisymmetry \citep[e.g.][]{
2015MNRAS.447..948C,2022MNRAS.512.1857S,2025A&A...699A.239F,2026ApJ..1002...71V}, or triaxiality \citep{2017MNRAS.466.4040F}. 
More recent works used discrete stellar velocities rather than binning the data, and combined the line-of-sight velocities, \vlos, obtained from spectroscopy with proper motions. This is usually not possible for extragalactic systems, where the kinematic information is often limited to line-of-sight data from the integrated light, and thus spatially binned rather than discrete. 

Nonetheless, besides a map of the integrated \vlos, more information can be extracted from integrated light data, such as maps of the velocity dispersion, \slos, or higher-order moments of the line-of-sight velocity distribution \citep[LOSVD, e.g.][]{2004PASP..116..138C}.  Even non-parametric LOSVDs have been extracted from optical spectroscopic data \citep{2019ApJ...887..195M,2021A&A...646A..31F}.
In combination with the stellar light distribution, such kinematic data can be used to infer the total mass distribution. Stellar dynamical modelling approaches make assumptions about the shape of the stellar systems (i.e. spherical, axisymmetric, or triaxial) and assume that the gravitational potential is static. Models based on the Jeans equations \citep{1922MNRAS..82..122J} require additional assumptions about the velocity structure. 
One of the commonly used approaches that does not make any such assumptions is based on \cite{1979ApJ...232..236S}; for example, in \cite{2008MNRAS.385..647V}, \cite{2013MNRAS.434.3174V} and \cite{2021MNRAS.500.1437N}. The codes have been improved and further developed in recent years \citep[e.g.][]{2019MNRAS.487.3776P,2020ascl.soft11007J,2022A&A...667A..51T}, and can even include barred structures \citep{2020ApJ...889...39V,2024MNRAS.534..861T,2025A&A...700A.249J}. This so-called orbit-based modelling has been validated on simulations several times \citep[e.g.][]{2019MNRAS.486.4753J,2020MNRAS.496.1579Z,2023MNRAS.519.2004N,2025A&A...700A.249J}.
Unbarred triaxial Schwarzschild models have been used to model the nuclear regions of galaxies and infer the central mass distribution and stellar orbital structure (\citealt{2013MNRAS.431.3364L,2017MNRAS.466.4040F,2019A&A...628A..92F,2021MNRAS.508.4786D,2023A&A...675A..18T,thatersubm,2026A&A...706A.373L}). 

The GC is an interesting object with which to test stellar dynamical models. However, several studies based on Jeans models underestimated \mbh \citep[e.g.][]{2009A&A...502...91S,2014A&A...570A...2F,2016ApJ...821...44F}. Also, triaxial \citep{2017MNRAS.466.4040F}  and spherical \citep{2019MNRAS.484.1166M} orbit-based models could barely recover the mass of \sgra. Their kinematic data covered only the inner region of the NSC. More recently, discrete axisymmetric Jeans models of \cite{2025A&A...699A.239F}, using more extended kinematic data and a precise stellar density distribution, recovered \mbh very accurately as (4.35$^{+0.24}_{-0.23})\cdot10^6\,\msun$. In this work, we test if we can also obtain the correct value for \mbh using triaxial orbit-based models and integrated light stellar kinematics, utilising the same density distribution and kinematic data that extend over a similar area as in \cite{2025A&A...699A.239F}. Our models also deliver the mass distribution and orbital structure of the NSC and inner NSD. We decompose the stellar orbits into dynamically cold (high angular momentum) and hot (low angular momentum) components. These shed light on the orbit distribution of the GC stellar structures.

This paper is organised as follows. We describe the spectroscopic data and kinematics in Sect.~\ref{sec:specdata}, and the orbit-based modelling approach in Sect.~\ref{sec:model}. We present our results in Sect.~\ref{sec:results}, discuss them in Sect.~\ref{sec:disc} and conclude in Sect.~\ref{sec:conclude}.


\section{Spectroscopic data}
\label{sec:specdata}
\subsection{Observations and data cube construction}
\label{sec:datacube}

\begin{figure*}
 \centering
 \includegraphics[width=18cm]{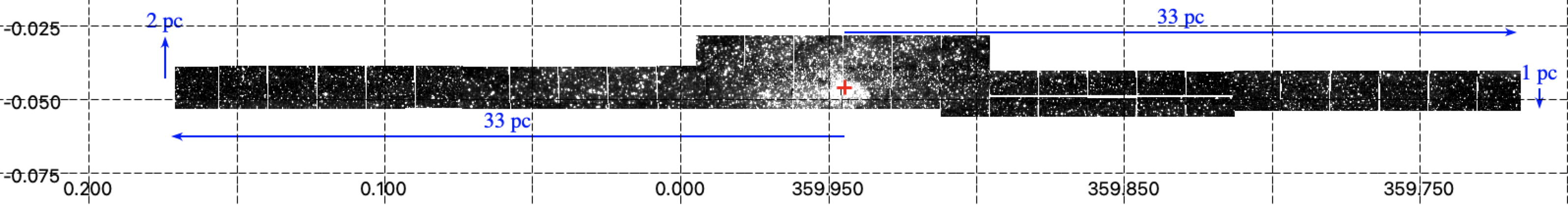}
 \caption{Spatial coverage of our F2 spectroscopic data. The data extend $\sim$66\,pc along the Galactic longitude $l$, centred on \sgra (marked as a red plus symbol), and  $\sim$1\,pc to the Galactic north and south, except for the centre region, which extends further to the Galactic north ($\sim$2\,pc). The image was constructed from the spectroscopic scans. We show the Galactic co-ordinate grid as dashed lines.}
 \label{fig:fov}
 \end{figure*}

The observations were taken on five nights (June 24, 25, 26, 27, and 29, 2015) with Flamingos-2 \citep[F2,][]{2004SPIE.5492.1196E} at the Gemini-South telescope. We observed in the near-infrared $K$ band, which is less affected by interstellar extinction ($A_{K_S}$$\approx$2.5\,mag, e.g. \citealt{2018A&A...610A..83N}) than optical wavelengths ($A_V$$\approx$30\,mag, \citealt{2011ApJ...737...73F}). 
The entire coverage of the spectroscopic data, meaning all fields combined, extends over $\sim$66\,pc along the Galactic longitude, $l$, with \sgra in the centre, and $\sim$2\,pc along the Galactic latitude, $b$ ($\sim$3\,pc in the inner $\left|l\right|\lesssim$8\,pc); see Fig.~\ref{fig:fov}.

The observations and basic data reduction are described in detail in \cite{2025A&A...696A.213F}. Here we just give a summary:
We observed five regions of the GC, which we labelled outer west, inner west, central, inner east, and outer east. These names are for Galactic co-ordinates and relative to \sgra. In each region, we observed 50 exposures (87 exposures in the central region) with a 6\arcmin\ long slit mask, and the telescope drifted by 1\arcsec\ per 300\,s exposure perpendicular to the slit mask. After 6 to 22 subsequent exposures, which we call a sequence, during which the telescope continuously drifted towards the Galactic South, we interrupted the series for sky and telluric calibration exposures. After these, we continued to scan the region to obtain 50 exposures in total. Due to these telescope offsets, in some cases, we have overlapping exposures or small gaps (see the horizontal white line in the inner west of Fig.~\ref{fig:fov}) between the observation sequences. We observed 20 sequences in total with the 6\arcmin\ long slit. The slit mask consists of six approximately 1\arcmin\ long slits or slitlets, aligned in a single row, with five small regions that were not cut to stabilise the mask. These regions cause small gaps in our data every 1\arcmin\ along $l$ (see vertical white lines in Fig.~\ref{fig:fov}). 

The data reduction includes dark subtraction, persistence subtraction, flat field division, cosmic ray removal, rectification, sky subtraction, and telluric correction. 
For each of the 20 sequences, divided into six slitlet regions, we constructed a stitched image. These 120 images were each cross-correlated with a Vista Variables in the Vía Láctea $K_S$ image \citep{2012A&A...537A.107S} to obtain their astrometric calibration. Figure~\ref{fig:fov} is a mosaic of these 120 images after astrometric calibration.

In \cite{2025A&A...696A.213F}, we extracted and analysed the spectra of the brightest stars, but there is also valuable information in the light of the fainter stars in the data. For this reason, we here use these data to construct data cubes of the unresolved light. 
In the data cubes, we masked all foreground and bright stars, as a few bright stars can dominate the light and outshine the unresolved stars \citep{2014A&A...570A...2F,2020AJ....160..146D}. We created masks as follows. We used the $JHK_S$ band photometry of the GALACTICNUCLEUS (GNS) survey catalogue \citep{2019A&A...631A..20N} to identify the approximate positions of stars on the slit. Starting with the brightest star per observing sequence, we fitted a Gaussian function to determine its exact location along the slit and then masked out a region of 7$\sigma$. 
For stars fainter than $K_S$\,=\,14\,mag, we considered only the primary exposure where the slit covers the star and no adjacent exposures. But brighter stars ($K_S$\,\textless\,14\,mag) can also contribute significantly to the exposure taken immediately before and after the one where the star is centred due to the seeing, and we also masked the flux of such stars in those exposures.
We created separate masks for foreground stars, which we identified using their $H-K_S$ colour. Foreground stars are less reddened than GC stars, and thus bluer. Our foreground star maps include all the stars with $H-K_S\leq$\,1.3\,mag \citep{2022MNRAS.513.5920F}, and for stars with unknown colour, where either the $H$ or $K_S$ band photometry are missing. This can be the case for bright stars that are saturated in the GNS survey.

As done for the stitched images, we combined subsequently taken exposures to stitched data cubes.  We used the previously created masks to remove the light from foreground stars, stars with unknown colour (as they may also be foreground stars), and bright stars ($K_{S,0}$\,=\,$K_S-A_{K_S}$\,\textless\,9.0\,mag). 
The brightness cut was chosen to be close to $K_{S,\rm cut}$ =\,11.5\,mag, to match the cut chosen by \cite{2014A&A...570A...2F}. However, instead of using a fixed value of $K_{S,\rm cut}$ for the observed $K_{S}$ band photometry, we fixed the value of the extinction corrected $K_{S}$ band photometry, $K_{S,0}$, to account for spatial variations in the extinction, $A_{K_S}$. We used the extinction map of \cite{2025A&A...696A.213F} and the respective mean extinction $A_{K_S}$ in the 120 image regions. 
For confirmed and possible foreground stars, we used an even stricter magnitude cut and masked them down to $K_S$\,\textless\,15\,mag.

We applied a wavelength calibration correction determined on the sky lines \citep[see][]{2025A&A...696A.213F} for each exposure.  We combined the 120 stitched data cubes such that the 50 exposures (86 in the centre) along latitude $b$ are combined to a single data cube, but keeping the separation into six  1\arcmin\ wide slitlets, using the code \textsc{ifsr\_mosaic.pro}\footnote{\url{https://github.com/drupke/ifsred}}. This results in 30 (5 regions $\times$ 6 slitlets) data cubes. We then rebinned the data cubes with \textsc{ifsr\_rebin.pro} to $\sim$1\arcsec $\cdot$ pixel$^{-1}$.

The median number of masked foreground stars per exposure and slitlet (covering $\sim$1\arcmin $\times$ 1\arcsec) is 0.8. The median number of masked bright stars is one. 
The final data cube contains only stars fainter than $K_S$\,$\approx$\,11.5\,mag.
According to the GNS catalogue, the median number of remaining stars with 11.5\,\textless\,$K_S$\,\textless\,15\,mag is 16 per exposure and slitlet. The minimum number is $\sim$8 stars in this magnitude range in the outer regions, and in the denser central region, we have a maximum of $\sim$70 stars.

\subsection{Spatial  binning}
\label{sec:vor}

We binned our 30 data cubes spatially before we measured the stellar kinematics. 
To ensure that we have a sufficiently large number of stars per bin, we used the S/N of the data cubes and applied the Voronoi binning code of \cite{2003MNRAS.342..345C}. We slightly modified the \textsc{python} code such that it does not start at the pixel with the highest flux but rather in the centre of the respective data cube. We required a target S/N of 75 per bin, leading to up to 45 bins in the central field data cubes but only 1 bin in some outer data cubes, where the stellar density is lower. In total, we have 197 Voronoi bins distributed over the entire region. For each Voronoi bin, we have a spectrum that contains the integrated flux of the unresolved stars in the region. 

Using a higher value for the target S/N per bin leads to coarser spatial binning in the centre of our field, averaging out spatial information. A lower target S/N produces a finer spatial binning, but at the cost of noisier kinematic maps. We tried different target S/N values (ranging from 45-95) and found that a target S/N of 75 leads to robust kinematic results without compromising valuable spatial information.

\begin{figure}
\resizebox{\hsize}{!}{\includegraphics{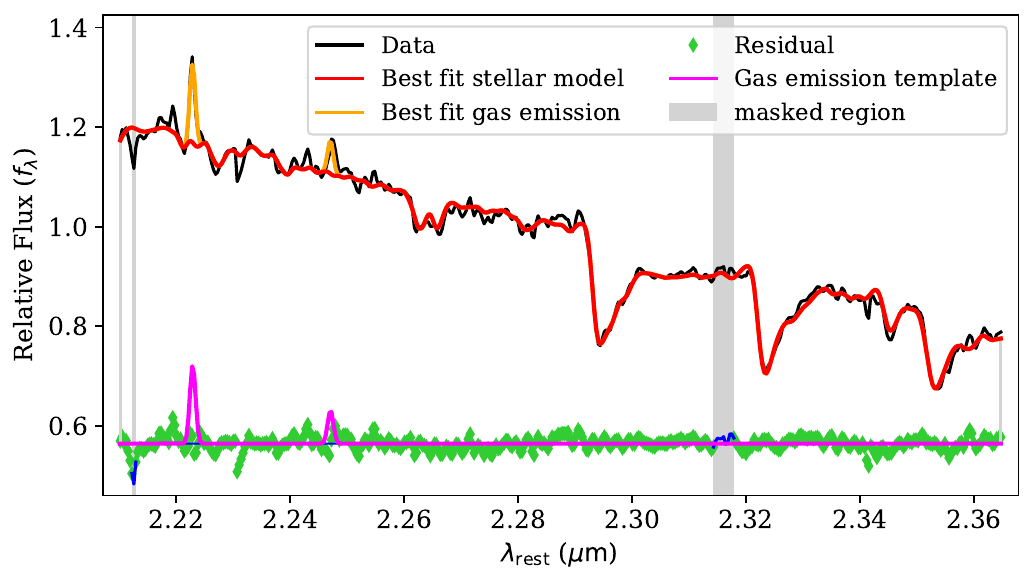}}
 \caption{Example of a stellar kinematic fit with \textsc{ppxf}. The black line shows the data, the red line the best-fit stellar model, and the orange and pink lines the best-fit gas emission. Green symbols denote residuals; the shaded grey regions were masked in the fit due to sky residuals or bad pixels. }
 \label{fig:spec}
\end{figure}

\subsection{Stellar kinematics}
\label{sec:kinfit}
We measured the stellar kinematics using the \textsc{python} code \textsc{pPXF}  
\citep{Cappellari2023} on the Voronoi binned spectra. This code requires template spectra, and we used the high-resolution spectral library of late-type stars by \cite{1996ApJS..107..312W}, convolved to the respective spectral resolution of the data, in the wavelength region of 2.21-2.365\,\micron. This wavelength region does not include the \ion{Na}{I} doublet, as we found systematic residuals in this region in some of the spectra, but includes the \ion{Ca}{I} triplet, and four CO transitions ($^{12}$CO (v=2-0) 2.2935\,\micron, $^{12}$CO (3-1) 2.3227\,\micron, $^{12}$CO (4-2) 2.3525\,\micron, and $^{13}$CO (2-0) 2.3448\,\micron). We fitted four Gauss-Hermite moments of the LOSVD. The first and second moments correspond to the velocity \vlos\ and velocity dispersion \slos\ in the limit where the Gauss-Hermite coefficients $h_3$ and $h_4$ are equal to zero. We used additive Legendre polynomials with degree=4 to correct the template continuum shape during the fit, as recommended for kinematic fits. 

The spectral resolution was measured in \cite{2025A&A...696A.213F} by fitting Gaussians to the sky emission lines. They found that the spectral resolution varies both as a function of wavelength and spatially for individual slitlets, with a maximum spectral resolution $R$=3\,400. For each slitlet, they derived a spectral resolution function,  which is a second-degree polynomial function of the wavelength, and we used these functions for the spectral analysis. 

It is necessary to include H$_2$ gas emission lines to fit the inner regions of our data. The gas emission originates from the circumnuclear disc (CND) of molecular gas in the inner $r$\,$\lesssim$\,3\,pc around \sgra. We included the transitions at 2.2235\,\micron\space (H$_2$ 1-0 S(0)) and 2.2477\,\micron\space (H$_2$ 2-1 S(1)), constrained to the same gas velocity and velocity dispersion. There are other H$_2$ transitions in the fitted wavelength region (e.g. 2.3556\,\micron\space H$_2$ 2-1 S(0)), but they are very weak, and we did not include them in the fit.

To reach our final kinematic results, we started with an initial fit, where we masked the wavelength region 2.3145--2.3175\,\micron, as this region can have residuals from the sky subtraction procedure. In a second fit, we masked in addition all pixels where the absolute value of the residual spectrum (= data -- best-fit) from the first fit exceeds 0.15. In a third fit, we set in addition the \textsc{pPXF} keyword \textsc{clean}, which does iterative sigma-clipping to remove unmasked bad pixels \citep{2002ApJ...578..787C}. We usually used the kinematic results from the third fit, unless more than 20 percent of all pixels were masked in this fit, then we used the result obtained by the second fit. The results are usually close, differences are only a few kilometres per second and within the measurement uncertainties. We show the example of a typical spectrum and the best-fit stellar and gas model in Fig.~\ref{fig:spec}. 
To obtain the kinematic errors, we performed Monte Carlo simulations. This was done by adding noise to the spectrum in 500 iterations and fitting the kinematics (with \textsc{clean}=False). We use the standard deviation of these fits as our kinematic uncertainties. 

We tested fits where we include the \ion{Na}{I} doublet, or to limit the fitting range to the first two CO lines ($^{12}$CO (v=2-0) 2.2935\,\micron, $^{12}$CO (3-1) 2.3227\,\micron). While these generally achieve similar results, our chosen wavelength range is better at dealing with spectra that are affected by sky residuals in the outer spatial regions of our data, leading to more robust results.

We also tried different template spectra for the fit. The X-SHOOTER spectral library (XSL) single stellar population (SSP) models \citep{2022A&A...661A..50V} have a sufficiently high spectral resolution and long wavelength range, but these templates had larger residuals at the \ion{Ca}{I} feature in comparison to the \cite{1996ApJS..107..312W} templates. This may be explained in the context of spectral index measurements of resolved stars. Stars in the GC have stronger \ion{Ca}{I} and \ion{Na}{I} EW values compared to Galactic disc stars \citep[e.g.][]{1996AJ....112.1988B,2017MNRAS.464..194F}. 
This trend also holds for integrated light spectra
\citep{2020AJ....160..146D}.
The \cite{1996ApJS..107..312W} stellar templates may have more flexibility in finding the best template than the XSL SSP models. 
The widely used SSP templates of \citet[][EMILES]{2016MNRAS.463.3409V} and \cite{2018ApJ...854..139C} have a lower spectral resolution than our data; hence, we chose not to use them.

\begin{figure*}
  \includegraphics[width=\columnwidth]{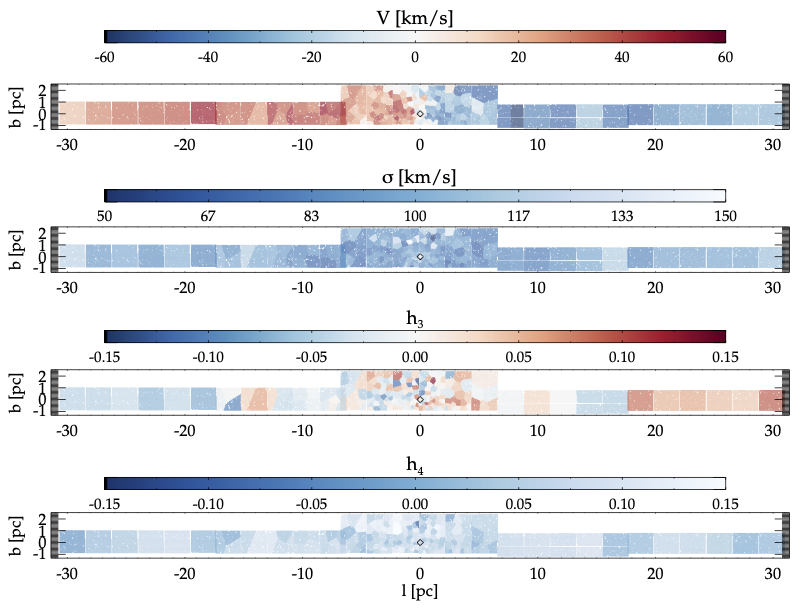}
  \includegraphics[width=\columnwidth]{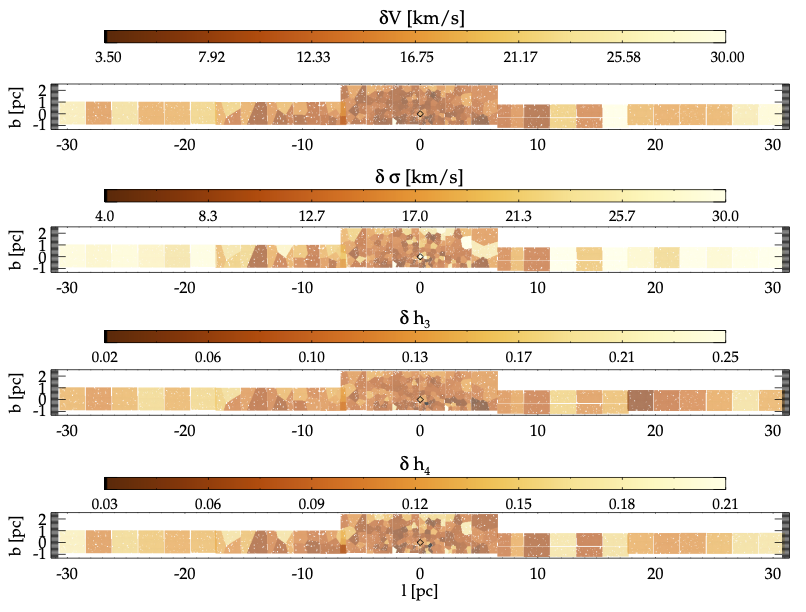}

  \caption{Stellar kinematic maps (left) after symmetrisation and their respective uncertainties (right). The plots show, from top to bottom, \vlos, \slos, $h_3$, and $h_4$.}
  \label{fig:kinmap}
\end{figure*}

We applied point-symmetrisation on our kinematic maps, using the \textsc{symmetrize\_velfield.py} code in the \textsc{plotbin python} package\footnote{\url{https://pypi.org/project/plotbin/}}. Symmetrising data reduces noise and removes systematic effects such as the systemic velocity in an extragalactic system \citep{2010MNRAS.401.1770V}, without causing significant biases on the modelling results \citep{2012ApJ...753...79W,2022MNRAS.509.5416T}.  
We do not alter the uncertainties of our maps. 
We show the point-symmetrised kinematic maps (left) and their uncertainties (right) in Fig.~\ref{fig:kinmap}.  The uncertainties are lower in the central region of the maps. Note the rotation seen in the \vlos map and the anti-correlation between \vlos and $h_3$.

\section{Orbit-based modelling}
\label{sec:model}
We used orbit-based modelling based on \cite{2008MNRAS.385..647V} with the tool \textsc{dynamite}  \citep[DYnamics, Age and Metallicity Indicators Tracing Evolution,][]{2020ascl.soft11007J,2022A&A...667A..51T}. The code finds the best combination of orbits in a given potential, and the best set of hyperparameters to describe the gravitational potential. The models are constrained by the four stellar kinematic maps shown in Fig.~\ref{fig:kinmap}. In this section, we describe how we model the gravitational potential and compute the orbit library.

\subsection{Gravitational potential} 
\label{sec:pot}
We assumed a gravitational potential that consists of a supermassive black hole (SMBH) and the stellar distribution. In a subset of models, we also included a spherical dark matter (DM) component. In our models, we assumed a galactocentric distance of 8.3\,kpc \citep{2022A&A...657L..12G}.

\subsubsection{Supermassive black hole}
\label{sec:mbh}
The SMBH was modelled as a Plummer sphere, with a scale radius of 0\farcs01. The scale radius is much smaller than the spatial resolution and pixel scale of our data; hence, the potential is a good approximation of a point mass. The SMBH mass, \mbh, is a free hyperparameter and was fit on a logarithmic scale. We restricted \mbh\ to 
{$10^{6.201}-10^{6.878}$\,\msun\ =\,(1.6 -- 7.6)\,$\cdot\,10^6$\,\msun}. The well-constrained value of \mbh = (4.3\,$\pm$\,0.012)\,$\cdot\,10^6$\,\msun \citep{2022A&A...657L..12G} was used as starting value.

\subsubsection{Stellar mass}
\label{sec:stellarmass}
We used the stellar number density map of \cite{2020A&A...634A..71G}, which traces red giant stars in the GC ($\sim$84.4\,pc $\times$ 21\,pc). The stars are in the extinction corrected magnitude range of 9.0\,$\leq\! K_{S,0}\! \leq$\,14.0\,mag. Our kinematic maps also trace red giant stars $K_{S,0}\! \geq$\,9.0\,mag (see Sect.~\ref{sec:datacube}). \cite{2025A&A...699A.239F} fitted a two-dimensional multi-Gaussian expansion \citep[MGE,][]{1994A&A...285..723E,2002MNRAS.333..400C} to the \cite{2020A&A...634A..71G} surface density map, and we used this MGE to model the surface stellar density \citep[see Table 1 in ][]{2025A&A...699A.239F}. The MGE was scaled to the 4.5\,\micron\ band surface light distribution of \cite{2017MNRAS.466.4040F} in the centre.

In our \textsc{dynamite} models, the deprojection of the surface stellar density MGE is determined by the space orientation angles ($\theta, \phi, \psi$), which can be converted analytically to three intrinsic shape parameters ($q$, $p$, $u$). These are ratios of the long, intermediate, and short axes $a$, $b$, and $c$ of a triaxial system, with $q=c/a$, $p=b/a$, and $u=a'/a$, where $a'$ denotes the length of the longest axis $a$ as projected on the sky. The flattest MGE component puts the strongest constraint on the deprojection, which means we have three free parameters (\qmin, \pmin, \umin) in our models \citep[see also][]{2018MNRAS.473.3000Z}. We limited \qmin to the range 0.1 -- 0.2999 (the upper limit is given by the MGE), \pmin to 0.6 -- 0.9999, and \umin to 0.8 -- 1.0, with starting points at \qmin=\,0.29, \pmin=\,0.94, and \umin=\,0.99. 

The deprojected stellar density distribution is multiplied by the dynamical mass-to-light ratio, \ups. We assumed that this factor is spatially constant (see \citealt{2025A&A...699A.239F}, who found that varying $\Upsilon$ is unnecessary), and used it to convert the stellar light density to a stellar mass density. We restricted its value to 0.1 -- 2.0, with a starting value at \ups\,=\,1.0. The hyperparameters \qmin, \pmin, \umin, and \ups thus describe the stellar gravitational potential. 

\subsubsection{Dark matter}
\label{sec:dm}
In a subset of models, we included a spherical DM component. The DM distribution is parameterised with a Navarro-Frenk-White \cite[NFW][]{1996ApJ...462..563N} profile with a mass-concentration from \cite{2014MNRAS.441.3359D}. Thus, we are left with only one parameter, the dark matter fraction, $f=M_{200}/M_*$, where $M_{200}$ is the mass enclosed within a radius $R_{200}$, the radius at which the average density is 200 times the critical density, and $M_*$ is the total stellar mass. The value of the hyperparameter $\log_{10} f$ is in the limits of -7 to +7 on a logarithmic scale, with a starting value at -3. 

\subsection{Constructing orbit solutions}
For each gravitational potential, constrained by the hyperparameters \qmin, \pmin, \umin, \mbh (and in a subset $f$), we numerically integrated a sample of orbits to form an orbit library, following the orbit-sampling scheme of \cite{2008MNRAS.385..647V}. The hyperparameter \ups only scales the potential and does not require a separate orbit integration. As also \mbh\ is scaled by \ups, the range of \mbh being modelled is 
{$\sim$(1.3 -- 10)\,$\cdot\,10^6$\,\msun}.

We have three orbit libraries per modelled potential: tube orbits, counter-rotating (CR) tube orbits, and box orbits. Each orbit library samples a combination of the three integrals of motion ($E$, $I_2$, and $I_3$, see \citealt{2008MNRAS.385..647V}), with a grid of $n_E\times n_{I_2} \times n_{I_3}$ = 35$\times$11$\times$11 orbits. The energy grid is sampled using the relation of energy to a circular orbit, and the radii of the circular orbits are in the range of 
10$^{0.5}$\,--\,10$^{3.86}$ arcsec, corresponding to 0.127\,--\,291.5\,pc. We note that the inner and outer radii correspond to $\sim$0.3$\,\times\,\sigma_{min}$ and $\sim$5\,$\times\,\sigma_{max}$, respectively, where $\sigma_{min}$ and $\sigma_{max}$ denote the $\sigma$ of the innermost and outermost MGE components in Sect.~\ref{sec:stellarmass}. 
We applied dithering with $3^3$, such that we actually have orbit bundles, and in total 3$\times$35$\times$11$\times$11$\times3^3$= 343\,035 orbits. 
The orbits were integrated for 200 periods, with a sampling of 50\,000 points per orbit. 

The orbit library weights were then fitted to reproduce the stellar LOSVD maps ($V$, $\sigma$, $h_3$, $h_4$) with the additional constraints that the deprojected 3D MGE stellar density distribution is reproduced within 1\%, and the observed 2D MGE surface density distribution within 2\% using a non-negative least squares (NNLS) minimisation. This is done for each set of hyperparameters. We computed the $\chi^2$ using the LOSVD maps, and found the best-fit model that minimises $\chi^2$. We refer to \cite{2008MNRAS.385..647V} and \cite{2018MNRAS.473.3000Z} for further details on the modelling and fitting procedure.

We sampled the parameter space with the \textsc{Dynamite} option \textsc{LegacyGridSearch}, which starts with a provided initial guess, then adds new models in later iterations. At each iteration, new models are seeded from existing models deemed acceptable based on a $\chi^2$ criterion. New models are seeded with gravitational potential parameters adjusted by some given step-size. The step-size decreases with later iterations to refine measurements. 
In total, we have computed \textgreater14\,000 models without DM in 16 iterations, and \textgreater3\,000 models with DM, in a narrower region of \mbh=(3.7-4.8)$\cdot 10^6$\,\msun. We illustrate the sampled grid in Figs.~\ref{fig:kinmapchi2} and \ref{fig:kinmapchi2_dm}.

For the 1$\sigma$ confidence level, we select all models within a tolerance $\Delta \chi^2$ of the best-fitting model. To determine $\Delta \chi^2$, one can use $\Delta \chi^2 = \sqrt{2\cdot n_{GH}\cdot N_{kin}}$, where $n_{GH}\,=\,4$ LOSVD parameters, and $N_{kin}$\,=\,197 Voronoi bins (see Sect.~\ref{sec:vor}). However, the validity of this level can be tested in a bootstrapping process \citep[as in][]{2024MNRAS.534..861T,2025A&A...700A.249J}, to account for the model's numerical noise and non-uniqueness of the orbit weight distribution. We perturbed our kinematic data 1\,000 times, assuming Gaussian error distributions given by the uncertainty of each of the 4\,$\times$\,197 measurements, and re-symmetrised the kinematic maps. Then we re-fit the orbit weights for these perturbed kinematic maps using the gravitational potential and orbit library of the best-fit model without DM. The resulting $\chi^2$ distribution has a Gaussian $\sigma$ that is $\sim$1.6\,$\times\,\sqrt{2\cdot n_{GH}\cdot N_{kin}}$, and we use this larger value as adjusted $\Delta \chi^2$ for the 1$\sigma$ confidence level of the modelling.

\section{Results}
\label{sec:results}

\begin{figure}
 
\includegraphics[angle=270,width=7cm]{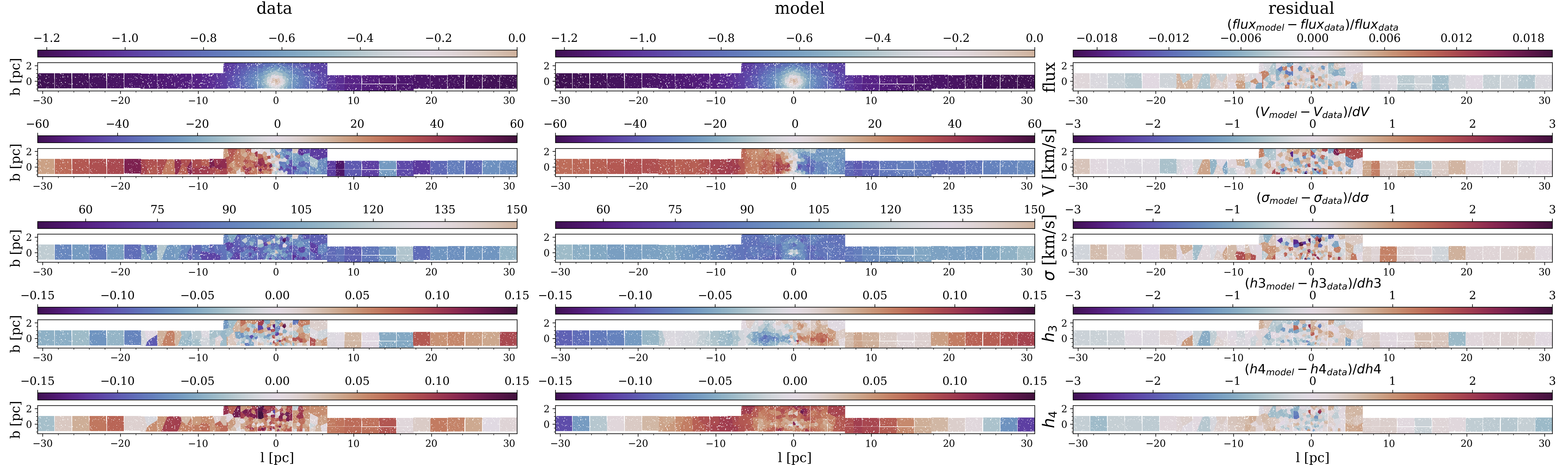}
      {\caption{Comparison of observations (left in rotated image), the best-fit model (middle, no DM), and the residuals (right). The rows show, from top to bottom: stellar flux, \vlos, \slos, $h_3$, and $h_4$.      }}
         \label{fig:kinmapsmodel}
   \end{figure}

Our best-fit results are listed in Table \ref{tab:bf}. 
The models with and without DM obtain the same best-fit results for \mbh, \ups, \qmin, \pmin, and \umin. The value of the DM fraction, $f$, is only poorly constrained. The best-fit model (without DM) flux and kinematics are compared to the data in Fig.~\ref{fig:kinmapsmodel}. The model and data agree very well, as indicated by the residual maps (right panel).

\begin{table}
	\centering
	\caption{Best-fit model parameters.}
	\label{tab:bf}
	\begin{tabular}{lrcccc} 
	\hline
      \noalign{\smallskip}
Parameter& Best fit      & 1$\sigma$ & 3$\sigma$ &$\Delta\chi^2$=5.9 & $\Delta\chi^2$=18.2\\
  \noalign{\smallskip}
  \hline
  \noalign{\smallskip}
  \mbh [$10^6 M_\odot$]&     4.3 & $^{+4.1}_{-2.9}$  &  $^{+5.5}_{-3.0}$ &  $^{+0.0}_{-2.2}$ &  $^{+0.0}_{-2.4}$   \\
  \noalign{\smallskip}
   \noalign{\smallskip}
  \ups [\msun/$L_\odot$]&    1.0  & $^{+0.9}_{-0.4}$ & $^{+1.0}_{-0.55}$ & $^{+0.3}_{-0.1}$ & $^{+0.6}_{-0.3}$  \\
  \noalign{\smallskip}
   \noalign{\smallskip}
  \qmin & 0.2999  & $^{+0.0}_{-0.1899}$  &$^{+0.0}_{-0.1899}$ & $^{+0.0}_{-0.0099}$  &$^{+0.0}_{-0.0799}$ \\
  \noalign{\smallskip}
   \noalign{\smallskip}
  \pmin& 0.92  & $^{+0.08}_{-0.18}$ & $^{+0.08}_{-0.23}$ & $^{+0.02}_{-0.10}$ & $^{+0.08}_{-0.10}$ \\
  \noalign{\smallskip}
   \noalign{\smallskip}
  \umin & 1.0  & $^{+0.0}_{-0.04}$ & $^{+0.0}_{-0.085}$ & $^{+0.0}_{-0.01}$ & $^{+0.0}_{-0.03}$  \\
  \noalign{\smallskip}
    \hline
  \noalign{\smallskip}
  $\log_{10} f$ &     $-5$ & $^{+10}_{-2}$  &  $^{+10}_{-2}$ &  $^{+6}_{-2}$ &  $^{+10}_{-2}$   \\
  \noalign{\smallskip}
 \hline
	\end{tabular}
    \tablefoot{\mbh, \ups, \qmin, \pmin, and \umin uncertainty ranges come from the models without DM. The best-fit values are identical in the models with DM. $\log_{10} f$ denotes $\log_{10}(M_{200}/M_*$). }
\end{table}

\subsection{Shape and orientation}
Our best-fit shape parameters \qmin and \umin prefer values near the edge of the model grid, at 0.2999 and 1.0, respectively. This indicates that the best-fit model is edge-on, and that the long axis is parallel to the Galactic plane. 
As the location of the Sun is relatively close to the Galactic mid-plane \citep[$\sim$21\,pc,][]{2019MNRAS.482.1417B}, this finding suggests that the GC region we model is aligned with the larger MW, and has the same inclination. 
The short axis is parallel to Galactic latitude, $b$, and the intermediate axis is along the line of sight towards \sgra. 

\begin{figure}
 \includegraphics[width=\columnwidth]{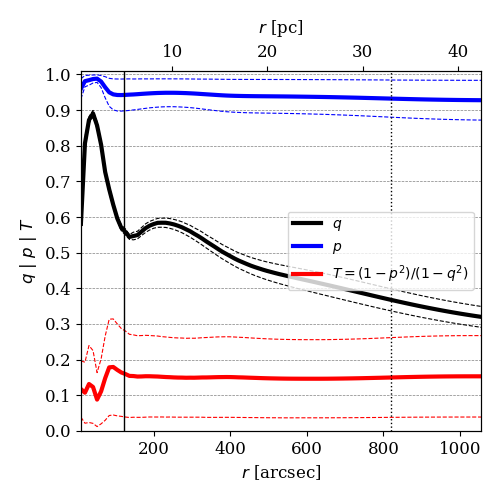}
  \caption{Intrinsic shape parameters and triaxiality as a function of radius $r$. The black lines denote $q=c/a$,  blue lines $p=b/a$, and red lines triaxiality $T=(1-p^2)/(1-q^2)$. Solid lines denote the best-fit model parameters, the dashed coloured lines their 1$\sigma$ uncertainties. The vertical solid line denotes 1~\re\ of the NSC, the dotted line the outer limit of the kinematic data. Horizontal lines mark steps of 0.1.}
  \label{fig:qpt}
\end{figure}

Our value of \pmin=\,0.92 deviates from 1.0, and thus from perfect oblate axisymmetry. Given the orientation of the axes,  $p$ indicates compression along the line of sight. We show its radial behaviour in Fig.~\ref{fig:qpt}. However, \pmin\ is closer to one than the result of \citet[\pmin=\,0.64]{2017MNRAS.466.4040F}, though the values agree within 1$\sigma$. 
The profile of $q$ indicates increasing flattening towards larger radii, which is expected as the inner NSC is less flattened than the surrounding NSD. 

The triaxiality parameter, T\,=\,$(1-p^2)/(1-q^2)$, quantifies the deviation from axisymmetry. A value of T\,=\,0 denotes oblate axisymmetry, T\,=\,1 prolate axisymmetry. As shown in Fig.~\ref{fig:qpt}, we obtain T in the range of roughly 0.03\,--\,0.27, with the best-fit value at T\,$\approx$\,0.15. This brings the models in the range of being at most mildly triaxial (T\,=\,0.1\,--\,0.3) according to the classification of \cite{2022ApJ...930..153S}.

\subsection{Mass distribution}

The value of \mbh matches the initial guess and true value of 4.3\,$\cdot\,10^6$\,\msun, though we have large uncertainties. Nonetheless, this confirms the validity of our models. Our result is an improvement on the triaxial Schwarzschild models of \cite{2017MNRAS.466.4040F}, who obtained \mbh\ =\,3.0\,$\cdot\,10^6$\,\msun, and the results agree within the uncertainties. 
Our value of \ups\,=\,1.0 also matches the expectations. We scaled the central surface density to the surface brightness profile of \citet[\ups=\,0.9]{2017MNRAS.466.4040F}, and our results for \ups are in agreement.

\begin{figure}
 \includegraphics[width=\columnwidth]{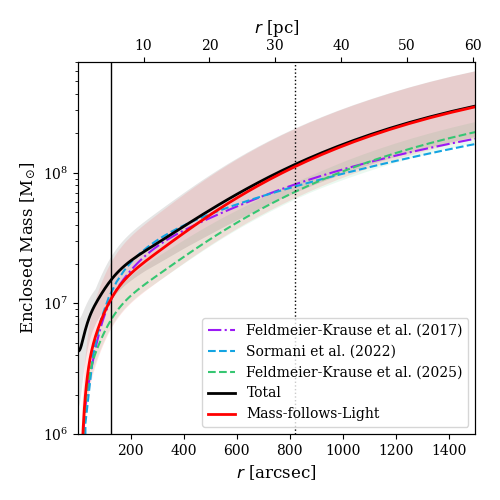}

  \caption{Total enclosed mass as a function of spherical deprojected radius. Shaded regions show 1$\sigma$ uncertainties. The vertical solid line denotes 1\,\re of the NSC, the dotted line the outer limit of the kinematic data.}
  \label{fig:mass}
\end{figure}

We show the mass distribution for the best-fit model without DM and the 1$\sigma$ uncertainty in Fig. \ref{fig:mass}, and list the enclosed mass at four different radii in Table \ref{tab:masses}. The best-fit model with DM results in a very similar total mass distribution as the model without DM. This is because the best-fit DM mass is more than five orders of magnitude lower than the total mass. 
For comparison, we also show the stellar mass profiles from other studies, and we will discuss the differences in Sect.~ \ref{sec:disc}.

\begin{table}
	\centering
	\caption{Enclosed total mass for the model with no DM, total mass, and DM mass for the model with DM at different deprojected spherical radii.}
	\label{tab:masses}
	\begin{tabular}{lcccc} 
	\hline
 Model& 5\,pc& 10\,pc& 20\,pc&30\,pc \\

 \hline
\noalign{\smallskip}  
Total mass [$10^7 M_\odot$]\\
\noalign{\smallskip}
\setlength\parindent{34pt} No DM&  1.5$^{+0.7}_{-0.4}$ & 2.5 $^{+1.7}_{-0.7}$ & 5.1 $^{+4.3}_{-1.8}$ &
9.8 $^{+8.7}_{-3.8}$ \\
\noalign{\smallskip}
\setlength\parindent{34pt} Including DM &1.5$^{+0.6}_{-0.4}$ & 2.5 $^{+1.3}_{-0.7}$ & 
5.1 $^{+3.5}_{-1.6}$ &
9.8$^{+6.8}_{-3.1}$ \\
\noalign{\smallskip}
\hline
\noalign{\smallskip}
DM mass [$10^3 M_\odot$]&  
0.3$^{+40.5}_{-0.3}$ & 
0.7 $^{+162.6}_{-0.7}$ & 
1.5 $^{+651.5}_{-1.5}$ &
2$^{+1468}_{-2}$ \\
\noalign{\smallskip}

 \hline
	\end{tabular}
\end{table}

\subsection{Orbit circularity distribution}
The circularity, $\lambda_z = \overline{L_z}/ (r\times \overline{V_c})$, indicates the orbit angular momentum, $\overline{L_z}$, around the short axis, normalised by the angular momentum of a circular orbit of the same binding energy. Thus, \lz=\,1 presents a strongly rotating and dynamically cold short-axis tube orbit, and \lz$=\,-1$ its retrograde counterpart, while \lz=\,0 is a dynamically hot box orbit. For each orbit, we computed the mean orbital radius, $\overline{r}$, from the radii stored at each equal time step during orbit integration. 
The distribution of \lz\ as a function of the mean orbital radius, $\overline{r}$, of a model allows us to assess which region is dominated by dynamical cold, warm, or hot components.

\begin{figure}
\includegraphics[width=\columnwidth]{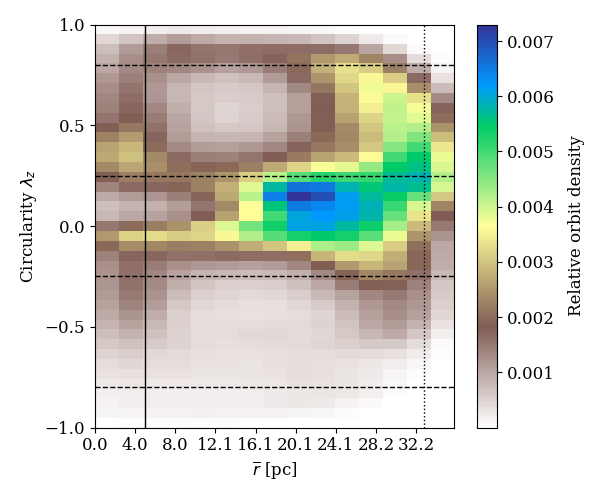}

  \caption{Orbit circularity distribution of $\lambda_z$ as a function of mean orbital radius $\overline{r}$, computed using over \textgreater800 models within 1$\sigma$. 
Colour indicates the orbit density in the phase space, horizontal dashed lines divide the orbits into cold ($\lambda_z$\textgreater0.8), warm  (0.25\textless$\lambda_z \leq$0.8), hot ($-0.25$\textless$\lambda_z\leq$0.25),  CR warm (--0.8\textless$\lambda_z\leq-0.2$), and CR cold ($\lambda_z\leq$-0.8) orbits, vertical lines are as in Fig. \ref{fig:qpt}.}
  \label{fig:orbitlinav}
\end{figure}

We show our circularity distribution  \lz\ in Fig.~\ref{fig:orbitlinav}. 
To create this plot, we used the orbit distributions of \textgreater800 models within the 1$\sigma$ uncertainty limit and computed the entropy-regularised Wasserstein barycentre \citep{2015SJSC...37A1111B,flamary2021pot,flamary2024pot}, which finds the transport-based middle of the 2-dimensional weight distributions. The weight distributions of the \textgreater800 models can differ by small shifts, and this method preserves the structure better, and produces sharper and physically more meaningful results than a simple average or median map. 

\begin{figure}
\includegraphics[width=\columnwidth]{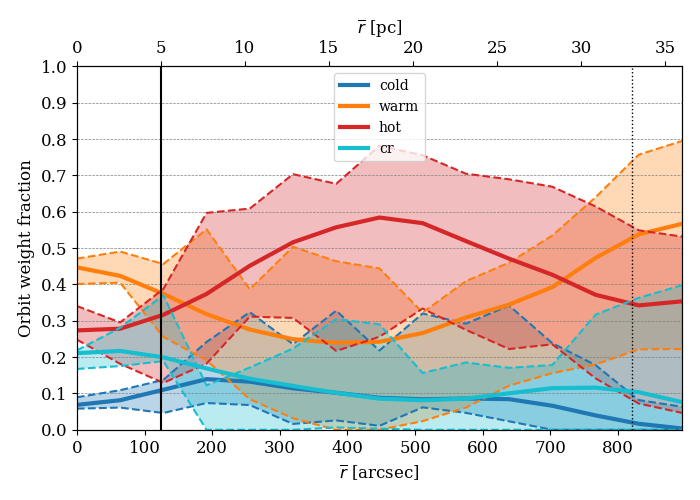}  \caption{Relative orbit weight profile as function of mean orbital radius $\overline{r}$, computed from the \lz\ distribution in Fig.~\ref{fig:orbitlinav}. 
The different colours denote different orbit types, the shaded regions the 1$\sigma$ percentiles of the 800 models. Blue denotes cold, orange warm,  red hot, and cyan CR orbits, vertical lines are as in Fig. \ref{fig:qpt}.}
  \label{fig:orbweight}
\end{figure}

We distinguish hot from warm orbits at $\lambda_z=\pm0.25$, and warm from cold orbits at $\lambda_z=\pm0.8$, using the same $\lambda_z$ cuts as \cite{2022ApJ...930..153S} and \cite{2023A&A...675A..18T}. 
To better understand how the relative contribution of different orbit types changes with $\overline{r}$, we show the relative weight of the cold, warm, hot, and CR orbits at a given mean orbital radius $\overline{r}$ in Fig. \ref{fig:orbweight}, the coloured bands show the 1$\sigma$ percentile ranges of \textgreater800 models within the 1$\sigma$ uncertainty limit.

In the region dominated by the NSC ($\overline{r}\lesssim$\,7\,pc), the models reveal that the largest contribution is from warm orbits, followed by hot, CR, and cold orbits. At a mean orbital radius of 5--15\,pc, that is 1--3\,\re of the NSC, the contribution of hot orbits increases from 0.3 to $\sim$0.6, whereas the weight fraction of warm orbits decreases from 0.4 to $\sim$0.25. These trends revert again at $\sim$20\,pc, when the warm orbits' weight fraction starts to increase, while the hot orbit weight fraction decreases.  The CR orbits decrease from $\sim$0.2 within the NSC to $\sim$0.1 in the inner NSD. Cold rotating orbits contribute $\sim$0.1 at these radii.

Our 1$\sigma$ models have large fractions of cold orbits and CR orbits in the outer NSD, though their exact location varies between individual models ($\overline{r}\sim$\,80--140\,pc). 
We note that the field of view (FOV) of the kinematic data extends only to $l$ = 33\,pc. Orbits with $\overline{r}$\textgreater33\,pc spend at least half of the time at intrinsic radii beyond the extent of the data. But stars on such orbits can cross our FOV if their projected radius is smaller. The presence of these cold orbits is constrained by our models, though the exact value of $\overline{r}$ is not. 

Another way to differentiate hot from cold orbits, but without distinction in co-rotating and CR orbits, is via $R_\text{max}$ vs $z_\text{max}$ plots. We show examples of such plots using the best-fit model in Appendix \ref{sec:rmaxzmax} 
to enable a comparison with \cite{2024A&A...685A..93N}, who integrated orbits of NSD stars in a fixed gravitational potential.

\subsection{Best-fit model orbit decomposition}
Here, we decompose the stellar orbits of our best-fit model. Using the orbit weight distribution and \lz, we show the spatial and kinematic distributions of different orbital types. This method has been applied to various galaxies in (\citealt{2018MNRAS.473.3000Z,2018MNRAS.479..945Z,2020MNRAS.491.1690J,2022ApJ...930..153S,2026arXiv260407297B}), but barely on NSCs (but see \citealt{2026A&A...706A.373L}).

\begin{figure}
\includegraphics[angle=270,width=7.9cm]{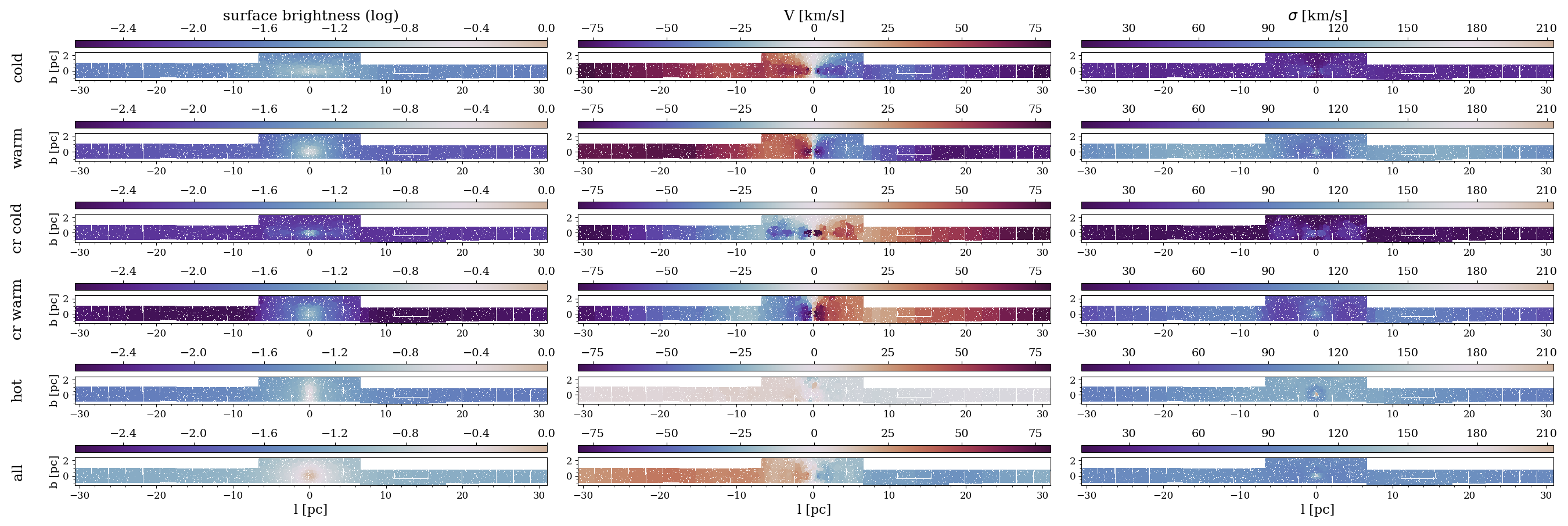}
      \caption{Orbit decomposition of the best-fit model in cold, warm, counter-rotating (cr) cold, cr warm, hot, and all orbits. The rows show, from left to right: stellar surface brightness, \vlos, \slos.      }
         \label{fig:decompmaps}
   \end{figure}

After grouping the orbits according to their \lz, we computed the surface brightness, \vlos, and \slos maps that the groups contribute to the total model. These maps are shown in Fig.~\ref{fig:decompmaps}. They take into account the contribution of all orbits that pass our FOV. 
We see that CR warm orbits have a steep decline in the surface brightness; they contribute most in the inner $\pm2$\,pc ($\sim$10\%), and little further out ($\lesssim$5\%). Cold orbits and hot orbits have the highest overall surface brightness contributions. Even though cold orbits appear sparse in the inner circularity plot (Figs.~\ref{fig:orbitlinav}--\ref{fig:orbweight}), they are important at large $\overline{r}$. These orbits cross the FOV, so that they contribute significantly to the surface brightness of the model. Except for the inner $\lesssim$2\,pc, where warm and hot orbits dominate, cold orbits have the highest relative weight to the surface brightness maps.

The \vlos maps reveal that the cold and warm orbits have \vlos values up to approximately $\pm$80\,\kms, but as there are also CR orbits and hot box orbits with lower absolute \vlos ($\lesssim\pm$10\,\kms), the \vlos value of all orbits combined adds up to only $\pm$40\,\kms. Interestingly, the CR cold orbits have the highest \vlos peaks in the inner $\sim$1\,pc with about $\pm$125\,\kms. 
The \slos maps show that cold orbits have lower \slos than warm orbits, and hot orbits have the highest \slos of all orbit types, as expected.

\section{Discussion}\label{sec:disc}

\subsection{Comparison to dynamical models in the literature}
Our best-fit model fits the data very well. Given our kinematic uncertainties and the flexibility of the Schwarzschild models, we found many models that fit similarly well. Subsequently, the parameter uncertainties are large, in particular, compared to \cite{2017MNRAS.466.4040F}, who used an older version of the \textsc{dynamite} code and constrained the models with different kinematic data and a different stellar MGE distribution. However, this is because we use a more conservative $\Delta\chi^2$ cut to define the parameter uncertainties than \cite{2017MNRAS.466.4040F} did, who used $\Delta \chi^2$=5.9 and 18.2 as 1$\sigma$ and 3$\sigma$ limits. With these bounds (see Table~\ref {tab:bf}), our uncertainties are much lower and similar to the uncertainties of \cite{2017MNRAS.466.4040F}. 

Nonetheless, we can constrain the shape and orbital distributions. The intrinsic shape parameters, in particular $q$, have only a rather small uncertainty (see  Fig.~\ref{fig:qpt}), and agree very well with the flattening of the stellar density measured by \cite{2020A&A...634A..71G}, that is 0.71$\pm$0.10 for the NSC  and 0.338$\pm$0.002 for the NSD. We find that the value of $p$, the intermediate-to-major axis ratio, is relatively constant with radius, and less compressed than found by \citet[\pmin=0.64]{2017MNRAS.466.4040F} in the inner $\sim$6\,pc. \cite{2017MNRAS.466.4040F} speculate that interstellar dust within the GC dominantly affects the stars further away along the LOS, and that the integrated light is biased to the near side of the GC. A possible consequence is that the GC appears compressed along the LOS, and the value of \pmin may be underestimated. Our FOV is larger and extends to less reddened regions, and our spectroscopic observations are deeper (we have integrated five times as long). These two factors may, to some extent, mitigate this LOS bias in our data. The anisotropy profiles (Fig.~\ref{fig:aniso}) show a remarkable agreement, with a minimum value within $r$\,$\lesssim$\,3\,pc, and close to isotropy at $r$\,$\gtrsim$\,4\,pc. 

We compare the stellar mass profiles from \citet[binned triaxial orbit-based models]{2017MNRAS.466.4040F}, \citet[discrete axisymmetric distribution function models]{2022MNRAS.512.1857S}, and \citet[discrete axisymmetric Jeans models]{2025A&A...699A.239F} in Fig.~\ref{fig:mass}. The stellar mass profiles agree mostly within the uncertainties, though they deviate beyond our uncertainties at $r$\,$\gtrsim$\,45\,pc, where we have no kinematic data.
The data of \cite{2017MNRAS.466.4040F} are limited to $l$\,$\lesssim$\,6\,pc. 
The models of \cite{2022MNRAS.512.1857S} are focused on the more extended NSD.  At $l$\,$\lesssim$\,20\,pc, their sample is rather small, as their data lies mostly at larger distances.  
We used the same light distribution as \cite{2025A&A...699A.239F}, but our mass distribution has larger uncertainties, by about a factor of 10. One reason for this is that Jeans models have less flexibility in comparison to the orbit-based modelling, and hence unrealistically small uncertainties. 
 \cite{2025A&A...699A.239F} had discrete data rather than binned data, which was more extended (covering a circle of $r$\,$\lesssim$\,9.5\,pc in addition to data in our FOV), and proper motions for $\sim$75\% of the 4\,600 stars with \vlos, leading to tighter constraints on the enclosed mass and mass-to-light ratio ($\Upsilon$=0.75$\pm$0.02).  
Further, the models of \cite{2025A&A...699A.239F} have a separate component for a background contribution of bar stars, which we do not have. This may bias our velocity dispersion to higher values, especially at the outer region of our kinematic data, as the bar contribution becomes more important at larger radii \citep{2025A&A...696A.213F}. This may cause a higher $\Upsilon$ and thus mass estimate.

Most of the previously mentioned works on the extended GC mass distribution neglected DM. \cite{2025A&A...699A.239F} included a DM component to a subset of their models. Like us, they obtained large uncertainties. An extrapolation of the DM volume density to $\gtrsim$100\,pc allows for a comparison with the simulations of \cite{2025arXiv250114868H} and the bar models of \cite{2017MNRAS.465.1621P}. While the \cite{2025A&A...699A.239F} models lie mostly above the values of \cite{2017MNRAS.465.1621P} and \cite{2025arXiv250114868H}, our DM volume density tends to be lower, yet both agree within the uncertainties. 
To be more quantitative, the \cite{2025A&A...699A.239F} DM mass contribution to the total enclosed mass lies at 6--25\% at 33\,pc, whereas we obtain only $\lesssim$0.8\%. The true value lies probably in between.

\subsection{Circularity distribution and orbit decomposition}
Our circularity distribution shows contributions from cold, warm, hot, and CR orbits. The circularity contains information on the origin of the stars. \cite{2023MNRAS.519.5202B} studied the circularity of the inner 500\,pc$^3$ of TNG50 galaxies and found that, at the stellar mass range of the MW (a few times 10$^{10}$\,\msun), stars that migrated to the galaxy centre have on average more rotational support (colder orbits, $\lambda_z$\,$\sim$\,0.5) compared to in situ stars, which show more random motion ($\lambda_z$\,$\sim$\,0.25). Ex situ stars, that is, stars accreted from galaxy mergers, are on average on hot orbits, but there is a large galaxy-to-galaxy variation. The circularity depends on the time of the merger and the orbit configuration between the host and merging satellite, and can result in CR orbits. \cite{2023MNRAS.519.5202B} also found that stars on colder orbits are, on average, younger, and the orbits become hotter due to dynamical heating over time (see also \citealt{2024A&A...692L..10B}).  The cold orbits we found at the outer NSD ($\overline{r}$\,$\gtrsim$\,80\,pc) may indeed be populated by younger stars, as the inside-out formation scenario for NSDs \citep{2020A&A...643A..65B,2023A&A...671L..10N} suggests. We note that most of the giant molecular gas structures of the central molecular zone are located within $\sim$100\,pc (except for the 1.3$^{\circ}$  cloud complex, \citealt{2016MNRAS.457.2675H}). 
However, as our data covers only a narrow range of latitude ($\lesssim$3\,pc), the circularity of cold orbits at $\overline{r}$\,$\gtrsim$80\,pc may be overestimated by our models, and the orbits may be warmer. Nonetheless, we note that \cite{2024A&A...685A..93N}, who integrated the orbits of \textgreater1\,100 stars at $l$$\lesssim$ 210 \,pc in the NSD, also found a large fraction of tube orbits ($\gtrsim$65\%).

In general, dynamical heating makes a dynamically cold, thin disc thicker and hotter. It drives the system towards isotropy, erasing signatures of the ex situ origin of stars. The heating mechanisms in the GC are two-body relaxation (among stars or with stellar remnants, \citealt{2005PhR...419...65A}), and massive perturbers (e.g. giant molecular clouds or star clusters, \citealt{2007ApJ...656..709P}). While the former is dominant in the inner parsec, the latter is dominant at $r\gtrsim$1.5 to $\sim$100\,pc \citep{2007ApJ...656..709P}. The Galactic bar further influences stellar orbits and redistributes angular momentum within the Galaxy \citep{1992MNRAS.259..328A}. 
Our finding that the inner regions are dominated by warm and hot orbits rather than cold orbits is in agreement with these regions being older and having experienced more dynamical heating. 

Counter-rotating orbits may trace infalling star clusters. \cite{2017MNRAS.464.3720T} studied the consecutive infall of star clusters to the GC and found that these can produce kinematic substructures if they come from random orbital directions. Though such signatures will probably be washed out by dynamical relaxation in a few gigayears \citep{2020ApJ...901L..29A}. \cite{2023A&A...674A..70I}  found that there may be up to 3--4 close passages of globular clusters with the GC per 1 Gyr. During these interactions, the star clusters can lose some of their stars, which then leave a dynamical imprint in the circularity distribution.

Some of the hot orbits in our model may be chaotic orbits. \cite{2025MNRAS.542..322P} computed the fraction of chaotic orbits among box orbits ($\lambda_z$=0) as a function of radius and NSC flattening, and found values up to 40\% for $q$=0.7. Assuming this fraction remains constant at $r$\,\textgreater\,10\,pc, we would have an overall fraction of $\sim$15\% chaotic orbits in our FOV. 
Galactic bar orbits are not explicitly included in our models. In the inner 100\,pc, we would expect some contribution from $x_1$, $x_2$, and $x_3$ orbits, extended along and perpendicular to the bar length. Though the overall contribution of the bar to the stellar surface density in the region of our data is $\lesssim$10\% at 30\,pc \citep{2025A&A...696A.213F}, and even less further in. 
\cite{2018NatAs...2..233Z} showed that triaxial orbit-based models that do not explicitly include a bar structure still reproduce the orbit distributions of simulated barred galaxies without strong biases. We conclude that this omission has no strong effect on our orbit distribution.

There have been speculations of a nuclear bar inside the GC \citep{2001A&A...379L..44A,2011A&A...534L..14G,2026A&A...709A..40F}, but no clear detection yet. Our models also show no sign of a nuclear bar. Bars produce diagonal lines in the \lz\ distribution plots \citep{2024MNRAS.534..861T}, which we did not obtain. 
We will investigate including bar orbits in orbit-based modelling of the GC in the future.

\section{Summary and conclusions}
\label{sec:conclude}
We have computed a large suite of triaxial orbit-based dynamical models of the GC, constrained by the integrated line-of-sight kinematics of red giant stars. The kinematic data extend out to 33\,pc from \sgra towards the Galactic east and west, and 1-2\,pc towards the north and south, and thus from the NSC to the inner part of the NSD. 

Our models recover the mass of the central black hole correctly, though with large uncertainties: \mbh=(4.3$^{+4.1}_{-2.9})\times10^6$\,\msun. Our stellar mass agrees with axisymmetric studies, though we tend to obtain higher masses at $r$\textgreater30\,pc. Including a dark matter component does not alter the other best-fit parameters, as we find only a very low DM contribution:  $\lesssim$1\% of the total mass at 33\,pc. 
Overall, our results are in good agreement with studies in the literature that use resolved kinematic data. This validates that orbit-based dynamical models with integrated light data, which are commonly used in extragalactic studies, produce robust results.

We find only mild triaxiality across the modelled region. The orbit circularity as a function of mean orbital radius shows mostly hot and warm orbits in the NSC and inner NSD. We detect a cold component at the outer part of the NSD ($\overline{r}\sim$80--160\,pc), though its exact location varies among 1 $\sigma$ models. This component has experienced less dynamical heating and may possibly be younger. We also detect some CR orbits, which may correspond to infalling star clusters.

To make the best use of orbit-based modelling in the GC, it would be ideal to fit the models to discrete velocity measurements rather than the integrated light maps and to include proper motion data. Such models have been attempted for axisymmetric \citep{2008ApJ...682..841C} and spherical systems \citep{2019MNRAS.484.1166M}, but with only a few applications.  In the future, the \textsc{dynamite} code will be extended to include proper motion and discrete data, and the GC is an ideal testbed for this code.

\begin{acknowledgements}
We thank the anonymous referee for their review and constructive comments.
A.F.K. acknowledges funding from the Austrian Science Fund (FWF) [grant DOI 10.55776/ESP542]. I.B. has received funding from the European Union’s Horizon 2020 research and innovation programme under the Marie Sklodowska-Curie Grant agreement ID n.o 101059532. This project was extended for 6
months by the Franziska Seidl Funding Program of the University of Vienna.  I.B. was supported by Fundação para a Ciência e a Tecnologia (FCT) through national funds under the research grant UID/04434/2025 (DOI 10.54499/UID/04434/2025).
We thank the Gemini Observatory staff for their support during the planning and execution of the observations and for their advice on data reduction. 
We thank David Rupke for providing a general-purpose library for IFU data cubes.
The computational results have been achieved using the Austrian Scientific Computing (ASC) infrastructure.
Based on observations obtained at the international Gemini Observatory, a program of NSF NOIRLab, which is managed by the Association of Universities for Research in Astronomy (AURA) under a cooperative agreement with the U.S. National Science Foundation on behalf of the Gemini Observatory partnership: the U.S. National Science Foundation (United States), National Research Council (Canada), Agencia Nacional de Investigaci\'{o}n y Desarrollo (Chile), Ministerio de Ciencia, Tecnolog\'{i}a e Innovaci\'{o}n (Argentina), Minist\'{e}rio da Ci\^{e}ncia, Tecnologia, Inova\c{c}\~{o}es e Comunica\c{c}\~{o}es (Brazil), and Korea Astronomy and Space Science Institute (Republic of Korea). Data were processed using the Gemini IRAF package. 
This research made use of Montage. It is funded by the National Science Foundation under Grant Number ACI-1440620, and was previously funded by the National Aeronautics and Space Administration's Earth Science Technology Office, Computation Technologies Project, under Cooperative Agreement Number NCC5-626 between NASA and the California Institute of Technology.
This research has made use of NASA's Astrophysics Data System;  the IPython package \citep{PER-GRA:2007};  Astropy, a community-developed core Python package for Astronomy \citep{2018AJ....156..123A, 2013A&A...558A..33A};  SciPy \citep{Virtanen_2020};  matplotlib, a Python library for publication quality graphics \citep{Hunter:2007};  NumPy \citep{harris2020array}; ds9, a tool for data visualization supported by the Chandra X-ray Science Center (CXC) and the High Energy Astrophysics Science Archive Center (HEASARC) with support from the JWST Mission office at the Space Telescope Science Institute for 3D visualization. 

\end{acknowledgements}


\bibliographystyle{aa} 
\bibliography{bibtex_2.bib} 

@ARTICLE{2010MNRAS.401.1770V,
       author = {{van den Bosch}, Remco C.~E. and {de Zeeuw}, P. Tim},
        title = "{Estimating black hole masses in triaxial galaxies}",
      journal = {\mnras},
         year = 2010,
        month = jan,
       volume = {401},
       number = {3},
        pages = {1770-1780},
          doi = {10.1111/j.1365-2966.2009.15832.x},
archivePrefix = {arXiv},
       eprint = {0910.0844},
 primaryClass = {astro-ph.CO},
       adsurl = {https://ui.adsabs.harvard.edu/abs/2010MNRAS.401.1770V}
}

@ARTICLE{2023A&A...671L..10N,
       author = {{Nogueras-Lara}, F. and {Schultheis}, M. and {Najarro}, F. and {Sormani}, M.~C. and {Gadotti}, D.~A. and {Rich}, R.~M.},
        title = "{Evidence of an age gradient along the line of sight in the nuclear stellar disc of the Milky Way}",
      journal = {\aap},
         year = 2023,
        month = mar,
       volume = {671},
          eid = {L10},
        pages = {L10},
          doi = {10.1051/0004-6361/202345941},
archivePrefix = {arXiv},
       eprint = {2302.02890},
 primaryClass = {astro-ph.GA},
       adsurl = {https://ui.adsabs.harvard.edu/abs/2023A&A...671L..10N}
}

@ARTICLE{2026ApJ..1002...71V,
       author = {{Vasiliev}, Eugene and {Feldmeier-Krause}, Anja and {Sormani}, Mattia C.},
        title = "{Distribution Function-based Modeling of Discrete Kinematic Datasets, in Application to the Milky Way Nuclear Star Cluster}",
      journal = {\apj},
         year = 2026,
        month = may,
       volume = {1002},
       number = {1},
          eid = {71},
        pages = {71},
          doi = {10.3847/1538-4357/ae5a30},
archivePrefix = {arXiv},
       eprint = {2603.29502},
 primaryClass = {astro-ph.GA},
       adsurl = {https://ui.adsabs.harvard.edu/abs/2026ApJ..1002...71V}
}

@ARTICLE{2022MNRAS.513.5920F,
       author = {{Feldmeier-Krause}, A.},
        title = "{Stellar populations in the transition region of nuclear star cluster and nuclear stellar disc}",
      journal = {\mnras},
         year = 2022,
        month = jul,
       volume = {513},
       number = {4},
        pages = {5920-5934},
          doi = {10.1093/mnras/stac1227},
archivePrefix = {arXiv},
       eprint = {2204.13723},
 primaryClass = {astro-ph.GA},
       adsurl = {https://ui.adsabs.harvard.edu/abs/2022MNRAS.513.5920F}
}

@ARTICLE{2022A&A...657L..12G,
       author = {{GRAVITY Collaboration} and {Abuter}, R. and {Aimar}, N. and {Amorim}, A. and {Ball}, J. and {Baub{\"o}ck}, M. and {Berger}, J.~P. and {Bonnet}, H. and {Bourdarot}, G. and {Brandner}, W. and {Cardoso}, V. and {Cl{\'e}net}, Y. and {Dallilar}, Y. and {Davies}, R. and {de Zeeuw}, P.~T. and {Dexter}, J. and {Drescher}, A. and {Eisenhauer}, F. and {F{\"o}rster Schreiber}, N.~M. and {Foschi}, A. and {Garcia}, P. and {Gao}, F. and {Gendron}, E. and {Genzel}, R. and {Gillessen}, S. and {Habibi}, M. and {Haubois}, X. and {Hei{\ss}el}, G. and {Henning}, T. and {Hippler}, S. and {Horrobin}, M. and {Jochum}, L. and {Jocou}, L. and {Kaufer}, A. and {Kervella}, P. and {Lacour}, S. and {Lapeyr{\`e}re}, V. and {Le Bouquin}, J. -B. and {L{\'e}na}, P. and {Lutz}, D. and {Ott}, T. and {Paumard}, T. and {Perraut}, K. and {Perrin}, G. and {Pfuhl}, O. and {Rabien}, S. and {Shangguan}, J. and {Shimizu}, T. and {Scheithauer}, S. and {Stadler}, J. and {Stephens}, A.~W. and {Straub}, O. and {Straubmeier}, C. and {Sturm}, E. and {Tacconi}, L.~J. and {Tristram}, K.~R.~W. and {Vincent}, F. and {von Fellenberg}, S. and {Widmann}, F. and {Wieprecht}, E. and {Wiezorrek}, E. and {Woillez}, J. and {Yazici}, S. and {Young}, A.},
        title = "{Mass distribution in the Galactic Center based on interferometric astrometry of multiple stellar orbits}",
      journal = {\aap},
         year = 2022,
        month = jan,
       volume = {657},
          eid = {L12},
        pages = {L12},
          doi = {10.1051/0004-6361/202142465},
archivePrefix = {arXiv},
       eprint = {2112.07478},
 primaryClass = {astro-ph.GA},
       adsurl = {https://ui.adsabs.harvard.edu/abs/2022A&A...657L..12G}
}

@ARTICLE{2022MNRAS.512.1857S,
       author = {{Sormani}, Mattia C. and {Sanders}, Jason L. and {Fritz}, Tobias K. and {Smith}, Leigh C. and {Gerhard}, Ortwin and {Sch{\"o}del}, Rainer and {Magorrian}, John and {Neumayer}, Nadine and {Nogueras-Lara}, Francisco and {Feldmeier-Krause}, Anja and {Mastrobuono-Battisti}, Alessandra and {Schultheis}, Mathias and {Shahzamanian}, Banafsheh and {Vasiliev}, Eugene and {Klessen}, Ralf S. and {Lucas}, Philip and {Minniti}, Dante},
        title = "{Self-consistent modelling of the Milky Way's nuclear stellar disc}",
      journal = {\mnras},
         year = 2022,
        month = may,
       volume = {512},
       number = {2},
        pages = {1857-1884},
          doi = {10.1093/mnras/stac639},
archivePrefix = {arXiv},
       eprint = {2111.12713},
 primaryClass = {astro-ph.GA},
       adsurl = {https://ui.adsabs.harvard.edu/abs/2022MNRAS.512.1857S}
}

@ARTICLE{2022A&A...661A..50V,
       author = {{Verro}, K. and {Trager}, S.~C. and {Peletier}, R.~F. and {Lan{\c{c}}on}, A. and {Arentsen}, A. and {Chen}, Y. -P. and {Coelho}, P.~R.~T. and {Dries}, M. and {Falc{\'o}n-Barroso}, J. and {Gonneau}, A. and {Lyubenova}, M. and {Martins}, L. and {Prugniel}, P. and {S{\'a}nchez-Bl{\'a}zquez}, P. and {Vazdekis}, A.},
        title = "{Modelling simple stellar populations in the near-ultraviolet to near-infrared with the X-shooter Spectral Library (XSL)}",
      journal = {\aap},
         year = 2022,
        month = may,
       volume = {661},
          eid = {A50},
        pages = {A50},
          doi = {10.1051/0004-6361/202142387},
archivePrefix = {arXiv},
       eprint = {2110.10190},
 primaryClass = {astro-ph.GA},
       adsurl = {https://ui.adsabs.harvard.edu/abs/2022A&A...661A..50V}
}

@ARTICLE{2004PASP..116..138C,
       author = {{Cappellari}, Michele and {Emsellem}, Eric},
        title = "{Parametric Recovery of Line-of-Sight Velocity Distributions from Absorption-Line Spectra of Galaxies via Penalized Likelihood}",
      journal = {\pasp},
         year = 2004,
        month = feb,
       volume = {116},
       number = {816},
        pages = {138-147},
          doi = {10.1086/381875},
archivePrefix = {arXiv},
       eprint = {astro-ph/0312201},
 primaryClass = {astro-ph},
       adsurl = {https://ui.adsabs.harvard.edu/abs/2004PASP..116..138C}
}

@ARTICLE{Cappellari2023,
    author = {{Cappellari}, M.},
    title = "{Full spectrum fitting with photometry in PPXF: stellar population
        versus dynamical masses, non-parametric star formation history and
        metallicity for 3200 LEGA-C galaxies at redshift $z\approx0.8$}",
    journal = {MNRAS},
    eprint = {2208.14974},
    year = 2023,
    volume = 526,
    pages = {3273-3300},
    doi = {10.1093/mnras/stad2597}
}

@ARTICLE{2018ApJ...854..139C,
       author = {{Conroy}, Charlie and {Villaume}, Alexa and {van Dokkum}, Pieter G. and {Lind}, Karin},
        title = "{Metal-rich, Metal-poor: Updated Stellar Population Models for Old Stellar Systems}",
      journal = {\apj},
         year = 2018,
        month = feb,
       volume = {854},
       number = {2},
          eid = {139},
        pages = {139},
          doi = {10.3847/1538-4357/aaab49},
archivePrefix = {arXiv},
       eprint = {1801.10185},
 primaryClass = {astro-ph.GA},
       adsurl = {https://ui.adsabs.harvard.edu/abs/2018ApJ...854..139C}
}

@ARTICLE{2016MNRAS.463.3409V,
       author = {{Vazdekis}, A. and {Koleva}, M. and {Ricciardelli}, E. and {R{\"o}ck}, B. and {Falc{\'o}n-Barroso}, J.},
        title = "{UV-extended E-MILES stellar population models: young components in massive early-type galaxies}",
      journal = {\mnras},
         year = 2016,
        month = dec,
       volume = {463},
       number = {4},
        pages = {3409-3436},
          doi = {10.1093/mnras/stw2231},
archivePrefix = {arXiv},
       eprint = {1612.01187},
 primaryClass = {astro-ph.GA},
       adsurl = {https://ui.adsabs.harvard.edu/abs/2016MNRAS.463.3409V}
}

@ARTICLE{2001A&A...379L..44A,
       author = {{Alard}, C.},
        title = "{Another bar in the Bulge}",
      journal = {\aap},
         year = 2001,
        month = dec,
       volume = {379},
        pages = {L44-L47},
          doi = {10.1051/0004-6361:20011487},
archivePrefix = {arXiv},
       eprint = {astro-ph/0110491},
 primaryClass = {astro-ph},
       adsurl = {https://ui.adsabs.harvard.edu/abs/2001A&A...379L..44A}
}

@ARTICLE{2011A&A...534L..14G,
       author = {{Gonzalez}, O.~A. and {Rejkuba}, M. and {Minniti}, D. and {Zoccali}, M. and {Valenti}, E. and {Saito}, R.~K.},
        title = "{The inner Galactic bar traced by the VVV survey}",
      journal = {\aap},
         year = 2011,
        month = oct,
       volume = {534},
          eid = {L14},
        pages = {L14},
          doi = {10.1051/0004-6361/201117959},
archivePrefix = {arXiv},
       eprint = {1110.0925},
 primaryClass = {astro-ph.GA},
       adsurl = {https://ui.adsabs.harvard.edu/abs/2011A&A...534L..14G}
}

@ARTICLE{2017MNRAS.464.3720T,
       author = {{Tsatsi}, Athanasia and {Mastrobuono-Battisti}, Alessandra and {van de Ven}, Glenn and {Perets}, Hagai B. and {Bianchini}, Paolo and {Neumayer}, Nadine},
        title = "{On the rotation of nuclear star clusters formed by cluster inspirals}",
      journal = {\mnras},
         year = 2017,
        month = jan,
       volume = {464},
       number = {3},
        pages = {3720-3727},
          doi = {10.1093/mnras/stw2593},
archivePrefix = {arXiv},
       eprint = {1610.01162},
 primaryClass = {astro-ph.GA},
       adsurl = {https://ui.adsabs.harvard.edu/abs/2017MNRAS.464.3720T}
}

@ARTICLE{2020A&A...643A..65B,
       author = {{Bittner}, Adrian and {S{\'a}nchez-Bl{\'a}zquez}, Patricia and {Gadotti}, Dimitri A. and {Neumann}, Justus and {Fragkoudi}, Francesca and {Coelho}, Paula and {de Lorenzo-C{\'a}ceres}, Adriana and {Falc{\'o}n-Barroso}, Jes{\'u}s and {Kim}, Taehyun and {Leaman}, Ryan and {Mart{\'\i}n-Navarro}, Ignacio and {M{\'e}ndez-Abreu}, Jairo and {P{\'e}rez}, Isabel and {Querejeta}, Miguel and {Seidel}, Marja K. and {van de Ven}, Glenn},
        title = "{Inside-out formation of nuclear discs and the absence of old central spheroids in barred galaxies of the TIMER survey}",
      journal = {\aap},
         year = 2020,
        month = nov,
       volume = {643},
          eid = {A65},
        pages = {A65},
          doi = {10.1051/0004-6361/202038450},
archivePrefix = {arXiv},
       eprint = {2009.01856},
 primaryClass = {astro-ph.GA},
       adsurl = {https://ui.adsabs.harvard.edu/abs/2020A&A...643A..65B}
}

@ARTICLE{2020ApJ...901L..29A,
       author = {{Arca Sedda}, Manuel and {Gualandris}, Alessia and {Do}, Tuan and {Feldmeier-Krause}, Anja and {Neumayer}, Nadine and {Erkal}, Denis},
        title = "{On the Origin of a Rotating Metal-poor Stellar Population in the Milky Way Nuclear Cluster}",
      journal = {\apjl},
         year = 2020,
        month = oct,
       volume = {901},
       number = {2},
          eid = {L29},
        pages = {L29},
          doi = {10.3847/2041-8213/abb245},
archivePrefix = {arXiv},
       eprint = {2009.02328},
 primaryClass = {astro-ph.GA},
       adsurl = {https://ui.adsabs.harvard.edu/abs/2020ApJ...901L..29A}
}

@ARTICLE{2020AJ....160..146D,
       author = {{Davidge}, T.~J.},
        title = "{The Near-infrared Spectrum of the Nuclear Star Cluster: Looking below the Tip of the Iceberg, and Comparisons with Extragalactic Nuclei}",
      journal = {\aj},
         year = 2020,
        month = sep,
       volume = {160},
       number = {3},
          eid = {146},
        pages = {146},
          doi = {10.3847/1538-3881/abab97},
archivePrefix = {arXiv},
       eprint = {2008.08061},
 primaryClass = {astro-ph.GA},
       adsurl = {https://ui.adsabs.harvard.edu/abs/2020AJ....160..146D}
}

@ARTICLE{2020A&ARv..28....4N,
       author = {{Neumayer}, Nadine and {Seth}, Anil and {B{\"o}ker}, Torsten},
        title = "{Nuclear star clusters}",
      journal = {\aapr},
         year = 2020,
        month = jul,
       volume = {28},
       number = {1},
          eid = {4},
        pages = {4},
          doi = {10.1007/s00159-020-00125-0},
archivePrefix = {arXiv},
       eprint = {2001.03626},
 primaryClass = {astro-ph.GA},
       adsurl = {https://ui.adsabs.harvard.edu/abs/2020A&ARv..28....4N}
}

@ARTICLE{2020A&A...634A..71G,
       author = {{Gallego-Cano}, E. and {Sch{\"o}del}, R. and {Nogueras-Lara}, F. and {Dong}, H. and {Shahzamanian}, B. and {Fritz}, T.~K. and {Gallego-Calvente}, A.~T. and {Neumayer}, N.},
        title = "{New constraints on the structure of the nuclear stellar cluster of the Milky Way from star counts and MIR imaging}",
      journal = {\aap},
         year = 2020,
        month = feb,
       volume = {634},
          eid = {A71},
        pages = {A71},
          doi = {10.1051/0004-6361/201935303},
archivePrefix = {arXiv},
       eprint = {2001.08182},
 primaryClass = {astro-ph.GA},
       adsurl = {https://ui.adsabs.harvard.edu/abs/2020A&A...634A..71G}
}

@ARTICLE{2019A&A...631A..20N,
       author = {{Nogueras-Lara}, F. and {Sch{\"o}del}, R. and {Gallego-Calvente}, A.~T. and {Dong}, H. and {Gallego-Cano}, E. and {Shahzamanian}, B. and {Girard}, J.~H.~V. and {Nishiyama}, S. and {Najarro}, F. and {Neumayer}, N.},
        title = "{GALACTICNUCLEUS: A high-angular-resolution JHK$_{s}$ imaging survey of the Galactic centre. II. First data release of the catalogue and the most detailed CMDs of the GC}",
      journal = {\aap},
         year = 2019,
        month = nov,
       volume = {631},
          eid = {A20},
        pages = {A20},
          doi = {10.1051/0004-6361/201936263},
archivePrefix = {arXiv},
       eprint = {1908.10366},
 primaryClass = {astro-ph.GA},
       adsurl = {https://ui.adsabs.harvard.edu/abs/2019A&A...631A..20N}
}

@ARTICLE{2019A&A...628A..92F,
       author = {{Fahrion}, Katja and {Lyubenova}, Mariya and {van de Ven}, Glenn and {Leaman}, Ryan and {Hilker}, Michael and {Mart{\'\i}n-Navarro}, Ignacio and {Zhu}, Ling and {Alfaro-Cuello}, Mayte and {Coccato}, Lodovico and {Corsini}, Enrico M. and {Falc{\'o}n-Barroso}, Jes{\'u}s and {Iodice}, Enrichetta and {McDermid}, Richard M. and {Sarzi}, Marc and {de Zeeuw}, Tim},
        title = "{Constraining nuclear star cluster formation using MUSE-AO observations of the early-type galaxy FCC 47}",
      journal = {\aap},
         year = 2019,
        month = aug,
       volume = {628},
          eid = {A92},
        pages = {A92},
          doi = {10.1051/0004-6361/201935832},
archivePrefix = {arXiv},
       eprint = {1907.01007},
 primaryClass = {astro-ph.GA},
       adsurl = {https://ui.adsabs.harvard.edu/abs/2019A&A...628A..92F}
}

@ARTICLE{2018A&A...610A..83N,
       author = {{Nogueras-Lara}, F. and {Gallego-Calvente}, A.~T. and {Dong}, H. and {Gallego-Cano}, E. and {Girard}, J.~H.~V. and {Hilker}, M. and {de Zeeuw}, P.~T. and {Feldmeier-Krause}, A. and {Nishiyama}, S. and {Najarro}, F. and {Neumayer}, N. and {Sch{\"o}del}, R.},
        title = "{GALACTICNUCLEUS: A high angular resolution JHK$_{s}$ imaging survey of the Galactic centre. I. Methodology, performance, and near-infrared extinction towards the Galactic centre}",
      journal = {\aap},
         year = 2018,
        month = mar,
       volume = {610},
          eid = {A83},
        pages = {A83},
          doi = {10.1051/0004-6361/201732002},
archivePrefix = {arXiv},
       eprint = {1709.09094},
 primaryClass = {astro-ph.GA},
       adsurl = {https://ui.adsabs.harvard.edu/abs/2018A&A...610A..83N}
}

@ARTICLE{2017ApJ...837...30G,
       author = {{Gillessen}, S. and {Plewa}, P.~M. and {Eisenhauer}, F. and {Sari}, R. and {Waisberg}, I. and {Habibi}, M. and {Pfuhl}, O. and {George}, E. and {Dexter}, J. and {von Fellenberg}, S. and {Ott}, T. and {Genzel}, R.},
        title = "{An Update on Monitoring Stellar Orbits in the Galactic Center}",
      journal = {\apj},
         year = 2017,
        month = mar,
       volume = {837},
       number = {1},
          eid = {30},
        pages = {30},
          doi = {10.3847/1538-4357/aa5c41},
archivePrefix = {arXiv},
       eprint = {1611.09144},
 primaryClass = {astro-ph.GA},
       adsurl = {https://ui.adsabs.harvard.edu/abs/2017ApJ...837...30G}
}

@ARTICLE{2023MNRAS.519.5202B,
       author = {{Boecker}, Alina and {Neumayer}, Nadine and {Pillepich}, Annalisa and {Frankel}, Neige and {Ramesh}, Rahul and {Leaman}, Ryan and {Hernquist}, Lars},
        title = "{The origin of stars in the inner 500 parsecs in TNG50 galaxies}",
      journal = {\mnras},
         year = 2023,
        month = mar,
       volume = {519},
       number = {4},
        pages = {5202-5235},
          doi = {10.1093/mnras/stac3759},
archivePrefix = {arXiv},
       eprint = {2301.11942},
 primaryClass = {astro-ph.GA},
       adsurl = {https://ui.adsabs.harvard.edu/abs/2023MNRAS.519.5202B}
}

@ARTICLE{1992MNRAS.259..328A,
       author = {{Athanassoula}, E.},
        title = "{Morphology of bar orbits.}",
      journal = {\mnras},
         year = 1992,
        month = nov,
       volume = {259},
        pages = {328-344},
          doi = {10.1093/mnras/259.2.328},
       adsurl = {https://ui.adsabs.harvard.edu/abs/1992MNRAS.259..328A}
}

@ARTICLE{2025MNRAS.542..322P,
       author = {{Penoyre}, Zephyr and {Rossi}, Elena Maria and {Stone}, Nicholas C.},
        title = "{Disruptions of stars and binary systems on chaotic orbits in an axisymmetric Milky Way centre}",
      journal = {\mnras},
         year = 2025,
        month = sep,
       volume = {542},
       number = {1},
        pages = {322-349},
          doi = {10.1093/mnras/staf1237},
archivePrefix = {arXiv},
       eprint = {2505.06344},
 primaryClass = {astro-ph.GA},
       adsurl = {https://ui.adsabs.harvard.edu/abs/2025MNRAS.542..322P}
}

@ARTICLE{2005PhR...419...65A,
       author = {{Alexander}, Tal},
        title = "{Stellar processes near the massive black hole in the Galactic center [review article]}",
      journal = {\physrep},
         year = 2005,
        month = nov,
       volume = {419},
       number = {2-3},
        pages = {65-142},
          doi = {10.1016/j.physrep.2005.08.002},
archivePrefix = {arXiv},
       eprint = {astro-ph/0508106},
 primaryClass = {astro-ph},
       adsurl = {https://ui.adsabs.harvard.edu/abs/2005PhR...419...65A}
}

@ARTICLE{2017MNRAS.466.4040F,
       author = {{Feldmeier-Krause}, A. and {Zhu}, L. and {Neumayer}, N. and {van de Ven}, G. and {de Zeeuw}, P.~T. and {Sch{\"o}del}, R.},
        title = "{Triaxial orbit-based modelling of the Milky Way Nuclear Star Cluster}",
      journal = {\mnras},
         year = 2017,
        month = apr,
       volume = {466},
       number = {4},
        pages = {4040-4052},
          doi = {10.1093/mnras/stw3377},
archivePrefix = {arXiv},
       eprint = {1701.01583},
 primaryClass = {astro-ph.GA},
       adsurl = {https://ui.adsabs.harvard.edu/abs/2017MNRAS.466.4040F}
}

@ARTICLE{2017MNRAS.464..194F,
       author = {{Feldmeier-Krause}, A. and {Kerzendorf}, W. and {Neumayer}, N. and {Sch{\"o}del}, R. and {Nogueras-Lara}, F. and {Do}, T. and {de Zeeuw}, P.~T. and {Kuntschner}, H.},
        title = "{KMOS view of the Galactic Centre - II. Metallicity distribution of late-type stars}",
      journal = {\mnras},
         year = 2017,
        month = jan,
       volume = {464},
       number = {1},
        pages = {194-209},
          doi = {10.1093/mnras/stw2339},
archivePrefix = {arXiv},
       eprint = {1610.01623},
 primaryClass = {astro-ph.GA},
       adsurl = {https://ui.adsabs.harvard.edu/abs/2017MNRAS.464..194F}
}

@ARTICLE{2016ApJ...830...17B,
       author = {{Boehle}, A. and {Ghez}, A.~M. and {Sch{\"o}del}, R. and {Meyer}, L. and {Yelda}, S. and {Albers}, S. and {Martinez}, G.~D. and {Becklin}, E.~E. and {Do}, T. and {Lu}, J.~R. and {Matthews}, K. and {Morris}, M.~R. and {Sitarski}, B. and {Witzel}, G.},
        title = "{An Improved Distance and Mass Estimate for Sgr A* from a Multistar Orbit Analysis}",
      journal = {\apj},
         year = 2016,
        month = oct,
       volume = {830},
       number = {1},
          eid = {17},
        pages = {17},
          doi = {10.3847/0004-637X/830/1/17},
archivePrefix = {arXiv},
       eprint = {1607.05726},
 primaryClass = {astro-ph.GA},
       adsurl = {https://ui.adsabs.harvard.edu/abs/2016ApJ...830...17B}
}

@ARTICLE{2016MNRAS.457.2122G,
       author = {{Georgiev}, Iskren Y. and {B{\"o}ker}, Torsten and {Leigh}, Nathan and {L{\"u}tzgendorf}, Nora and {Neumayer}, Nadine},
        title = "{Masses and scaling relations for nuclear star clusters, and their co-existence with central black holes}",
      journal = {\mnras},
         year = 2016,
        month = apr,
       volume = {457},
       number = {2},
        pages = {2122-2138},
          doi = {10.1093/mnras/stw093},
archivePrefix = {arXiv},
       eprint = {1601.02613},
 primaryClass = {astro-ph.GA},
       adsurl = {https://ui.adsabs.harvard.edu/abs/2016MNRAS.457.2122G}
}

@ARTICLE{2016ApJ...821...44F,
       author = {{Fritz}, T.~K. and {Chatzopoulos}, S. and {Gerhard}, O. and {Gillessen}, S. and {Genzel}, R. and {Pfuhl}, O. and {Tacchella}, S. and {Eisenhauer}, F. and {Ott}, T.},
        title = "{The Nuclear Cluster of the Milky Way: Total Mass and Luminosity}",
      journal = {\apj},
         year = 2016,
        month = apr,
       volume = {821},
       number = {1},
          eid = {44},
        pages = {44},
          doi = {10.3847/0004-637X/821/1/44},
archivePrefix = {arXiv},
       eprint = {1406.7568},
 primaryClass = {astro-ph.GA},
       adsurl = {https://ui.adsabs.harvard.edu/abs/2016ApJ...821...44F}
}

@ARTICLE{2014A&A...570A...2F,
       author = {{Feldmeier}, A. and {Neumayer}, N. and {Seth}, A. and {Sch{\"o}del}, R. and {L{\"u}tzgendorf}, N. and {de Zeeuw}, P.~T. and {Kissler-Patig}, M. and {Nishiyama}, S. and {Walcher}, C.~J.},
        title = "{Large scale kinematics and dynamical modelling of the Milky Way nuclear star cluster}",
      journal = {\aap},
         year = 2014,
        month = oct,
       volume = {570},
          eid = {A2},
        pages = {A2},
          doi = {10.1051/0004-6361/201423777},
archivePrefix = {arXiv},
       eprint = {1406.2849},
 primaryClass = {astro-ph.GA},
       adsurl = {https://ui.adsabs.harvard.edu/abs/2014A&A...570A...2F}
}

@ARTICLE{2012A&A...537A.107S,
       author = {{Saito}, R.~K. and {Hempel}, M. and {Minniti}, D. and {Lucas}, P.~W. and {Rejkuba}, M. and {Toledo}, I. and {Gonzalez}, O.~A. and {Alonso-Garc{\'\i}a}, J. and {Irwin}, M.~J. and {Gonzalez-Solares}, E. and {Hodgkin}, S.~T. and {Lewis}, J.~R. and {Cross}, N. and {Ivanov}, V.~D. and {Kerins}, E. and {Emerson}, J.~P. and {Soto}, M. and {Am{\^o}res}, E.~B. and {Gurovich}, S. and {D{\'e}k{\'a}ny}, I. and {Angeloni}, R. and {Beamin}, J.~C. and {Catelan}, M. and {Padilla}, N. and {Zoccali}, M. and {Pietrukowicz}, P. and {Moni Bidin}, C. and {Mauro}, F. and {Geisler}, D. and {Folkes}, S.~L. and {Sale}, S.~E. and {Borissova}, J. and {Kurtev}, R. and {Ahumada}, A.~V. and {Alonso}, M.~V. and {Adamson}, A. and {Arias}, J.~I. and {Bandyopadhyay}, R.~M. and {Barb{\'a}}, R.~H. and {Barbuy}, B. and {Baume}, G.~L. and {Bedin}, L.~R. and {Bellini}, A. and {Benjamin}, R. and {Bica}, E. and {Bonatto}, C. and {Bronfman}, L. and {Carraro}, G. and {Chen{\`e}}, A.~N. and {Clari{\'a}}, J.~J. and {Clarke}, J.~R.~A. and {Contreras}, C. and {Corvill{\'o}n}, A. and {de Grijs}, R. and {Dias}, B. and {Drew}, J.~E. and {Fari{\~n}a}, C. and {Feinstein}, C. and {Fern{\'a}ndez-Laj{\'u}s}, E. and {Gamen}, R.~C. and {Gieren}, W. and {Goldman}, B. and {Gonz{\'a}lez-Fern{\'a}ndez}, C. and {Grand}, R.~J.~J. and {Gunthardt}, G. and {Hambly}, N.~C. and {Hanson}, M.~M. and {He{\l}miniak}, K.~G. and {Hoare}, M.~G. and {Huckvale}, L. and {Jord{\'a}n}, A. and {Kinemuchi}, K. and {Longmore}, A. and {L{\'o}pez-Corredoira}, M. and {Maccarone}, T. and {Majaess}, D. and {Mart{\'\i}n}, E.~L. and {Masetti}, N. and {Mennickent}, R.~E. and {Mirabel}, I.~F. and {Monaco}, L. and {Morelli}, L. and {Motta}, V. and {Palma}, T. and {Parisi}, M.~C. and {Parker}, Q. and {Pe{\~n}aloza}, F. and {Pietrzy{\'n}ski}, G. and {Pignata}, G. and {Popescu}, B. and {Read}, M.~A. and {Rojas}, A. and {Roman-Lopes}, A. and {Ruiz}, M.~T. and {Saviane}, I. and {Schreiber}, M.~R. and {Schr{\"o}der}, A.~C. and {Sharma}, S. and {Smith}, M.~D. and {Sodr{\'e}}, L. and {Stead}, J. and {Stephens}, A.~W. and {Tamura}, M. and {Tappert}, C. and {Thompson}, M.~A. and {Valenti}, E. and {Vanzi}, L. and {Walton}, N.~A. and {Weidmann}, W. and {Zijlstra}, A.},
        title = "{VVV DR1: The first data release of the Milky Way bulge and southern plane from the near-infrared ESO public survey VISTA variables in the V{\'\i}a L{\'a}ctea}",
      journal = {\aap},
         year = 2012,
        month = jan,
       volume = {537},
          eid = {A107},
        pages = {A107},
          doi = {10.1051/0004-6361/201118407},
archivePrefix = {arXiv},
       eprint = {1111.5511},
 primaryClass = {astro-ph.GA},
       adsurl = {https://ui.adsabs.harvard.edu/abs/2012A&A...537A.107S}
}

@ARTICLE{2011ApJ...737...73F,
       author = {{Fritz}, T.~K. and {Gillessen}, S. and {Dodds-Eden}, K. and {Lutz}, D. and {Genzel}, R. and {Raab}, W. and {Ott}, T. and {Pfuhl}, O. and {Eisenhauer}, F. and {Yusef-Zadeh}, F.},
        title = "{Line Derived Infrared Extinction toward the Galactic Center}",
      journal = {\apj},
         year = 2011,
        month = aug,
       volume = {737},
       number = {2},
          eid = {73},
        pages = {73},
          doi = {10.1088/0004-637X/737/2/73},
archivePrefix = {arXiv},
       eprint = {1105.2822},
 primaryClass = {astro-ph.GA},
       adsurl = {https://ui.adsabs.harvard.edu/abs/2011ApJ...737...73F}
}

@ARTICLE{2009A&A...502...91S,
       author = {{Sch{\"o}del}, R. and {Merritt}, D. and {Eckart}, A.},
        title = "{The nuclear star cluster of the Milky Way: proper motions and mass}",
      journal = {\aap},
         year = 2009,
        month = jul,
       volume = {502},
       number = {1},
        pages = {91-111},
          doi = {10.1051/0004-6361/200810922},
archivePrefix = {arXiv},
       eprint = {0902.3892},
 primaryClass = {astro-ph.GA},
       adsurl = {https://ui.adsabs.harvard.edu/abs/2009A&A...502...91S}
}

@INPROCEEDINGS{2004SPIE.5492.1196E,
       author = {{Eikenberry}, Stephen S. and {Elston}, Richard and {Raines}, S. Nicholas and {Julian}, Jeff and {Corley}, Richard J. and {Hanna}, Kevin and {Hon}, David and {Julian}, Roger and {Rashkin}, David and {Leckie}, Brian and {Gardhouse}, W. Rusty and {Fletcher}, Murray and {Dunn}, Jennifer and {Wooff}, Robert},
        title = "{FLAMINGOS-2: the facility near-infrared wide-field imager and multi-object spectrograph for Gemini}",
    booktitle = {Ground-based Instrumentation for Astronomy},
         year = 2004,
       editor = {{Moorwood}, Alan F.~M. and {Iye}, Masanori},
       series = {Society of Photo-Optical Instrumentation Engineers (SPIE) Conference Series},
       volume = {5492},
        month = sep,
        pages = {1196-1207},
          doi = {10.1117/12.549796},
       adsurl = {https://ui.adsabs.harvard.edu/abs/2004SPIE.5492.1196E}
}

@ARTICLE{2017MNRAS.465.1621P,
       author = {{Portail}, Matthieu and {Gerhard}, Ortwin and {Wegg}, Christopher and {Ness}, Melissa},
        title = "{Dynamical modelling of the galactic bulge and bar: the Milky Way's pattern speed, stellar and dark matter mass distribution}",
      journal = {\mnras},
         year = 2017,
        month = feb,
       volume = {465},
       number = {2},
        pages = {1621-1644},
          doi = {10.1093/mnras/stw2819},
archivePrefix = {arXiv},
       eprint = {1608.07954},
 primaryClass = {astro-ph.GA},
       adsurl = {https://ui.adsabs.harvard.edu/abs/2017MNRAS.465.1621P}
}

@ARTICLE{2002ApJ...578..787C,
       author = {{Cappellari}, M. and {Verolme}, E.~K. and {van der Marel}, R.~P. and {Verdoes Kleijn}, G.~A. and {Illingworth}, G.~D. and {Franx}, M. and {Carollo}, C.~M. and {de Zeeuw}, P.~T.},
        title = "{The Counterrotating Core and the Black Hole Mass of IC 1459}",
      journal = {\apj},
         year = 2002,
        month = oct,
       volume = {578},
       number = {2},
        pages = {787-805},
          doi = {10.1086/342653},
archivePrefix = {arXiv},
       eprint = {astro-ph/0202155},
 primaryClass = {astro-ph},
       adsurl = {https://ui.adsabs.harvard.edu/abs/2002ApJ...578..787C}
}

@ARTICLE{2003MNRAS.342..345C,
       author = {{Cappellari}, Michele and {Copin}, Yannick},
        title = "{Adaptive spatial binning of integral-field spectroscopic data using Voronoi tessellations}",
      journal = {\mnras},
         year = 2003,
        month = jun,
       volume = {342},
       number = {2},
        pages = {345-354},
          doi = {10.1046/j.1365-8711.2003.06541.x},
archivePrefix = {arXiv},
       eprint = {astro-ph/0302262},
 primaryClass = {astro-ph},
       adsurl = {https://ui.adsabs.harvard.edu/abs/2003MNRAS.342..345C}
}

@ARTICLE{2002A&A...384..112L,
       author = {{Launhardt}, R. and {Zylka}, R. and {Mezger}, P.~G.},
        title = "{The nuclear bulge of the Galaxy. III. Large-scale physical characteristics of stars and interstellar matter}",
      journal = {\aap},
         year = 2002,
        month = mar,
       volume = {384},
        pages = {112-139},
          doi = {10.1051/0004-6361:20020017},
archivePrefix = {arXiv},
       eprint = {astro-ph/0201294},
 primaryClass = {astro-ph},
       adsurl = {https://ui.adsabs.harvard.edu/abs/2002A&A...384..112L}
}

@ARTICLE{1999A&A...348..768P,
       author = {{Philipp}, S. and {Zylka}, R. and {Mezger}, P.~G. and {Duschl}, W.~J. and {Herbst}, T. and {Tuffs}, R.~J.},
        title = "{The nuclear bulge. I. K band observations of the central 30 PC}",
      journal = {\aap},
         year = 1999,
        month = aug,
       volume = {348},
        pages = {768-782},
       adsurl = {https://ui.adsabs.harvard.edu/abs/1999A&A...348..768P}
}

@ARTICLE{1996AJ....112.1988B,
       author = {{Blum}, R.~D. and {Sellgren}, K. and {Depoy}, D.~L.},
        title = "{Really Cool Stars at the Galactic Center}",
      journal = {\aj},
         year = 1996,
        month = nov,
       volume = {112},
        pages = {1988},
          doi = {10.1086/118157},
archivePrefix = {arXiv},
       eprint = {astro-ph/9608107},
 primaryClass = {astro-ph},
       adsurl = {https://ui.adsabs.harvard.edu/abs/1996AJ....112.1988B}
}

@ARTICLE{1996ApJS..107..312W,
       author = {{Wallace}, L. and {Hinkle}, K.},
        title = "{High-Resolution Spectra of Ordinary Cool Stars in the K Band}",
      journal = {\apjs},
         year = 1996,
        month = nov,
       volume = {107},
        pages = {312},
          doi = {10.1086/192367},
       adsurl = {https://ui.adsabs.harvard.edu/abs/1996ApJS..107..312W}
}

@ARTICLE{2023A&A...674A..70I,
       author = {{Ishchenko}, Maryna and {Sobolenko}, Margaryta and {Kuvatova}, Dana and {Panamarev}, Taras and {Berczik}, Peter},
        title = "{Milky Way globular clusters on cosmological timescales. II. Interaction with the Galactic centre}",
      journal = {\aap},
         year = 2023,
        month = jun,
       volume = {674},
          eid = {A70},
        pages = {A70},
          doi = {10.1051/0004-6361/202245753},
archivePrefix = {arXiv},
       eprint = {2304.02311},
 primaryClass = {astro-ph.GA},
       adsurl = {https://ui.adsabs.harvard.edu/abs/2023A&A...674A..70I}
}

@ARTICLE{2007ApJ...656..709P,
       author = {{Perets}, Hagai B. and {Hopman}, Clovis and {Alexander}, Tal},
        title = "{Massive Perturber-driven Interactions between Stars and a Massive Black Hole}",
      journal = {\apj},
         year = 2007,
        month = feb,
       volume = {656},
       number = {2},
        pages = {709-720},
          doi = {10.1086/510377},
archivePrefix = {arXiv},
       eprint = {astro-ph/0606443},
 primaryClass = {astro-ph},
       adsurl = {https://ui.adsabs.harvard.edu/abs/2007ApJ...656..709P}
}

@ARTICLE{2025A&A...696A.213F,
       author = {{Feldmeier-Krause}, A. and {Neumayer}, N. and {Seth}, A. and {van de Ven}, G. and {Hilker}, M. and {Kissler-Patig}, M. and {Kuntschner}, H. and {L{\"u}tzgendorf}, N. and {Mastrobuono-Battisti}, A. and {Nogueras-Lara}, F. and {Perets}, H.~B. and {Sch{\"o}del}, R. and {Zocchi}, A.},
        title = "{A spectroscopic map of the Galactic centre: Observations and resolved stars}",
      journal = {\aap},
         year = 2025,
        month = apr,
       volume = {696},
          eid = {A213},
        pages = {A213},
          doi = {10.1051/0004-6361/202453414},
archivePrefix = {arXiv},
       eprint = {2503.11856},
 primaryClass = {astro-ph.GA},
       adsurl = {https://ui.adsabs.harvard.edu/abs/2025A&A...696A.213F}
}

@ARTICLE{2025arXiv250114868H,
       author = {{Hussein}, Abdelaziz and {Necib}, Lina and {Kaplinghat}, Manoj and {Kim}, Stacy Y. and {Wetzel}, Andrew and {Read}, Justin I. and {Rey}, Martin P. and {Agertz}, Oscar},
        title = "{Theoretical Predictions for the Inner Dark Matter Distribution in the Milky Way Informed by Simulations}",
      journal = {arXiv e-prints},
         year = 2025,
        month = jan,
          eid = {arXiv:2501.14868},
        pages = {arXiv:2501.14868},
          doi = {10.48550/arXiv.2501.14868},
archivePrefix = {arXiv},
       eprint = {2501.14868},
 primaryClass = {hep-ph},
       adsurl = {https://ui.adsabs.harvard.edu/abs/2025arXiv250114868H}
}

@ARTICLE{2024MNRAS.534..861T,
       author = {{Tahmasebzadeh}, Behzad and {Zhu}, Ling and {Shen}, Juntai and {Gadotti}, Dimitri A. and {Valluri}, Monica and {Thater}, Sabine and {van de Ven}, Glenn and {Jin}, Yunpeng and {Gerhard}, Ortwin and {Erwin}, Peter and {Jethwa}, Prashin and {Zocchi}, Alice and {Lilley}, Edward J. and {Fragkoudi}, Francesca and {de Lorenzo-C{\'a}ceres}, Adriana and {M{\'e}ndez-Abreu}, Jairo and {Neumann}, Justus and {Guo}, Rui},
        title = "{Schwarzschild modelling of barred s0 galaxy NGC 4371}",
      journal = {\mnras},
         year = 2024,
        month = oct,
       volume = {534},
       number = {1},
        pages = {861-882},
          doi = {10.1093/mnras/stae2109},
archivePrefix = {arXiv},
       eprint = {2310.00497},
 primaryClass = {astro-ph.GA},
       adsurl = {https://ui.adsabs.harvard.edu/abs/2024MNRAS.534..861T}
}

@ARTICLE{2016MNRAS.457.2675H,
       author = {{Henshaw}, J.~D. and {Longmore}, S.~N. and {Kruijssen}, J.~M.~D. and {Davies}, B. and {Bally}, J. and {Barnes}, A. and {Battersby}, C. and {Burton}, M. and {Cunningham}, M.~R. and {Dale}, J.~E. and {Ginsburg}, A. and {Immer}, K. and {Jones}, P.~A. and {Kendrew}, S. and {Mills}, E.~A.~C. and {Molinari}, S. and {Moore}, T.~J.~T. and {Ott}, J. and {Pillai}, T. and {Rathborne}, J. and {Schilke}, P. and {Schmiedeke}, A. and {Testi}, L. and {Walker}, D. and {Walsh}, A. and {Zhang}, Q.},
        title = "{Molecular gas kinematics within the central 250 pc of the Milky Way}",
      journal = {\mnras},
         year = 2016,
        month = apr,
       volume = {457},
       number = {3},
        pages = {2675-2702},
          doi = {10.1093/mnras/stw121},
archivePrefix = {arXiv},
       eprint = {1601.03732},
 primaryClass = {astro-ph.GA},
       adsurl = {https://ui.adsabs.harvard.edu/abs/2016MNRAS.457.2675H}
}

@ARTICLE{2013ARA&A..51..511K,
       author = {{Kormendy}, John and {Ho}, Luis C.},
        title = "{Coevolution (Or Not) of Supermassive Black Holes and Host Galaxies}",
      journal = {\araa},
         year = 2013,
        month = aug,
       volume = {51},
       number = {1},
        pages = {511-653},
          doi = {10.1146/annurev-astro-082708-101811},
archivePrefix = {arXiv},
       eprint = {1304.7762},
 primaryClass = {astro-ph.CO},
       adsurl = {https://ui.adsabs.harvard.edu/abs/2013ARA&A..51..511K}
}

@ARTICLE{2021MNRAS.508.4786D,
       author = {{den Brok}, Mark and {Krajnovi{\'c}}, Davor and {Emsellem}, Eric and {Brinchmann}, Jarle and {Maseda}, Michael},
        title = "{Dynamical modelling of the twisted galaxy PGC 046832}",
      journal = {\mnras},
         year = 2021,
        month = dec,
       volume = {508},
       number = {4},
        pages = {4786-4805},
          doi = {10.1093/mnras/stab2852},
archivePrefix = {arXiv},
       eprint = {2109.14640},
 primaryClass = {astro-ph.GA},
       adsurl = {https://ui.adsabs.harvard.edu/abs/2021MNRAS.508.4786D}
}

@ARTICLE{2024A&A...685A..93N,
       author = {{Nieuwmunster}, N. and {Schultheis}, M. and {Sormani}, M. and {Fragkoudi}, F. and {Nogueras-Lara}, F. and {Sch{\"o}del}, R. and {McMillan}, P. and {Smith}, L.~C. and {Sanders}, J.~L.},
        title = "{Orbital analysis of stars in the nuclear stellar disc of the Milky Way}",
      journal = {\aap},
         year = 2024,
        month = may,
       volume = {685},
          eid = {A93},
        pages = {A93},
          doi = {10.1051/0004-6361/202349000},
archivePrefix = {arXiv},
       eprint = {2403.00761},
 primaryClass = {astro-ph.GA},
       adsurl = {https://ui.adsabs.harvard.edu/abs/2024A&A...685A..93N}
}

@ARTICLE{2019ApJ...887..195M,
       author = {{Mehrgan}, Kianusch and {Thomas}, Jens and {Saglia}, Roberto and {Mazzalay}, Ximena and {Erwin}, Peter and {Bender}, Ralf and {Kluge}, Matthias and {Fabricius}, Maximilian},
        title = "{A 40 Billion Solar-mass Black Hole in the Extreme Core of Holm 15A, the Central Galaxy of Abell 85}",
      journal = {\apj},
         year = 2019,
        month = dec,
       volume = {887},
       number = {2},
          eid = {195},
        pages = {195},
          doi = {10.3847/1538-4357/ab5856},
archivePrefix = {arXiv},
       eprint = {1907.10608},
 primaryClass = {astro-ph.GA},
       adsurl = {https://ui.adsabs.harvard.edu/abs/2019ApJ...887..195M}
}

@ARTICLE{2021MNRAS.500.1437N,
       author = {{Neureiter}, B. and {Thomas}, J. and {Saglia}, R. and {Bender}, R. and {Finozzi}, F. and {Krukau}, A. and {Naab}, T. and {Rantala}, A. and {Frigo}, M.},
        title = "{SMART: a new implementation of Schwarzschild's Orbit Superposition technique for triaxial galaxies and its application to an N-body merger simulation}",
      journal = {\mnras},
         year = 2021,
        month = jan,
       volume = {500},
       number = {1},
        pages = {1437-1465},
          doi = {10.1093/mnras/staa3014},
archivePrefix = {arXiv},
       eprint = {2009.08979},
 primaryClass = {astro-ph.GA},
       adsurl = {https://ui.adsabs.harvard.edu/abs/2021MNRAS.500.1437N}
}

@ARTICLE{2023MNRAS.519.2004N,
       author = {{Neureiter}, B. and {de Nicola}, S. and {Thomas}, J. and {Saglia}, R. and {Bender}, R. and {Rantala}, A.},
        title = "{Accuracy and precision of triaxial orbit models I: SMBH mass, stellar mass, and dark-matter halo}",
      journal = {\mnras},
         year = 2023,
        month = feb,
       volume = {519},
       number = {2},
        pages = {2004-2016},
          doi = {10.1093/mnras/stac3652},
archivePrefix = {arXiv},
       eprint = {2212.06173},
 primaryClass = {astro-ph.GA},
       adsurl = {https://ui.adsabs.harvard.edu/abs/2023MNRAS.519.2004N}
}

@ARTICLE{2022ApJ...930..153S,
       author = {{Santucci}, Giulia and {Brough}, Sarah and {van de Sande}, Jesse and {McDermid}, Richard M. and {van de Ven}, Glenn and {Zhu}, Ling and {D'Eugenio}, Francesco and {Bland-Hawthorn}, Joss and {Barsanti}, Stefania and {Bryant}, Julia J. and {Croom}, Scott M. and {Davies}, Roger L. and {Green}, Andrew W. and {Lawrence}, Jon S. and {Lorente}, Nuria P.~F. and {Owers}, Matt S. and {Poci}, Adriano and {Richards}, Samuel N. and {Thater}, Sabine and {Yi}, Sukyoung},
        title = "{The SAMI Galaxy Survey: The Internal Orbital Structure and Mass Distribution of Passive Galaxies from Triaxial Orbit-superposition Schwarzschild Models}",
      journal = {\apj},
         year = 2022,
        month = may,
       volume = {930},
       number = {2},
          eid = {153},
        pages = {153},
          doi = {10.3847/1538-4357/ac5bd5},
archivePrefix = {arXiv},
       eprint = {2203.03648},
 primaryClass = {astro-ph.GA},
       adsurl = {https://ui.adsabs.harvard.edu/abs/2022ApJ...930..153S}
}

@ARTICLE{2021A&A...646A..31F,
       author = {{Falc{\'o}n-Barroso}, J. and {Martig}, M.},
        title = "{BAYES-LOSVD: A Bayesian framework for non-parametric extraction of the line-of-sight velocity distribution of galaxies}",
      journal = {\aap},
         year = 2021,
        month = feb,
       volume = {646},
          eid = {A31},
        pages = {A31},
          doi = {10.1051/0004-6361/202039624},
archivePrefix = {arXiv},
       eprint = {2011.12023},
 primaryClass = {astro-ph.GA},
       adsurl = {https://ui.adsabs.harvard.edu/abs/2021A&A...646A..31F}
}

@software{2020ascl.soft11007J,
       author = {{Jethwa}, Prashin and {Thater}, Sabine and {Maindl}, Thomas and {Van de Ven}, Glenn},
        title = "{DYNAMITE: DYnamics, Age and Metallicity Indicators Tracing Evolution}",
 howpublished = {Astrophysics Source Code Library, record ascl:2011.007},
         year = 2020,
        month = nov,
          eid = {ascl:2011.007},
       adsurl = {https://ui.adsabs.harvard.edu/abs/2020ascl.soft11007J}
}

@ARTICLE{2020MNRAS.496.1579Z,
       author = {{Zhu}, Ling and {van de Ven}, Glenn and {Leaman}, Ryan and {Grand}, Robert J.~J. and {Falc{\'o}n-Barroso}, Jes{\'u}s and {Jethwa}, Prashin and {Watkins}, Laura L. and {Mao}, Shude and {Poci}, Adriano and {McDermid}, Richard M. and {Nelson}, Dylan},
        title = "{Disentangling the formation history of galaxies via population-orbit superposition: method validation}",
      journal = {\mnras},
         year = 2020,
        month = aug,
       volume = {496},
       number = {2},
        pages = {1579-1597},
          doi = {10.1093/mnras/staa1584},
archivePrefix = {arXiv},
       eprint = {2003.05561},
 primaryClass = {astro-ph.GA},
       adsurl = {https://ui.adsabs.harvard.edu/abs/2020MNRAS.496.1579Z}
}

@ARTICLE{2019MNRAS.487.3776P,
       author = {{Poci}, Adriano and {McDermid}, Richard M. and {Zhu}, Ling and {van de Ven}, Glenn},
        title = "{Combining stellar populations with orbit-superposition dynamical modelling: the formation history of the lenticular galaxy NGC 3115}",
      journal = {\mnras},
         year = 2019,
        month = aug,
       volume = {487},
       number = {3},
        pages = {3776-3796},
          doi = {10.1093/mnras/stz1154},
archivePrefix = {arXiv},
       eprint = {1904.11605},
 primaryClass = {astro-ph.GA},
       adsurl = {https://ui.adsabs.harvard.edu/abs/2019MNRAS.487.3776P}
}

@ARTICLE{2019MNRAS.484.1166M,
       author = {{Magorrian}, John},
        title = "{Orbit-superposition models of discrete, incomplete stellar kinematics: application to the Galactic centre}",
      journal = {\mnras},
         year = 2019,
        month = mar,
       volume = {484},
       number = {1},
        pages = {1166-1181},
          doi = {10.1093/mnras/stz037},
archivePrefix = {arXiv},
       eprint = {1809.05946},
 primaryClass = {astro-ph.GA},
       adsurl = {https://ui.adsabs.harvard.edu/abs/2019MNRAS.484.1166M}
}

@ARTICLE{2018NatAs...2..233Z,
       author = {{Zhu}, Ling and {van de Ven}, Glenn and {van den Bosch}, Remco and {Rix}, Hans-Walter and {Lyubenova}, Mariya and {Falc{\'o}n-Barroso}, Jes{\'u}s and {Martig}, Marie and {Mao}, Shude and {Xu}, Dandan and {Jin}, Yunpeng and {Obreja}, Aura and {Grand}, Robert J.~J. and {Dutton}, Aaron A. and {Macci{\`o}}, Andrea V. and {G{\'o}mez}, Facundo A. and {Walcher}, Jakob C. and {Garc{\'\i}a-Benito}, Rub{\'e}n and {Zibetti}, Stefano and {S{\'a}nchez}, Sebastian F.},
        title = "{The stellar orbit distribution in present-day galaxies inferred from the CALIFA survey}",
      journal = {Nature Astronomy},
         year = 2018,
        month = jan,
       volume = {2},
        pages = {233-238},
          doi = {10.1038/s41550-017-0348-1},
archivePrefix = {arXiv},
       eprint = {1711.06728},
 primaryClass = {astro-ph.GA},
       adsurl = {https://ui.adsabs.harvard.edu/abs/2018NatAs...2..233Z}
}

@ARTICLE{2026A&A...709A..40F,
       author = {{Fiteni}, K. and {Li}, X. and {Sormani}, M.~C. and {Debattista}, V.~P. and {Vasini}, A. and {Nogueras-Lara}, F. and {Sanders}, J.~L. and {Deg}, N. and {Schultheis}, M. and {Donati}, M. and {Feng}, Z.-X.},
        title = "{Kinematic diagnostics for non-axisymmetry in the Milky Way's nuclear stellar disc}",
      journal = {\aap},
         year = 2026,
        month = apr,
       volume = {709},
          eid = {A40},
        pages = {A40},
          doi = {10.1051/0004-6361/202558739},
archivePrefix = {arXiv},
       eprint = {2603.18738},
 primaryClass = {astro-ph.GA},
       adsurl = {https://ui.adsabs.harvard.edu/abs/2026A&A...709A..40F}
}

@ARTICLE{2026arXiv260407297B,
       author = {{Breda}, Iris and {van de Ven}, Glenn and {Thater}, Sabine and {Falc{\'o}n-Barroso}, J. and {Jethwa}, Prashin and {Onodera}, Masato and {Schaye}, Joop and {Brinchmann}, Jarle and {Ziegler}, Bodo and {Mauro}, Federica},
        title = "{Tracing the dynamical and structural complexity of spiral galaxy centres}",
      journal = {\aap\ accepted},
         year = 2026,
        month = apr,
          eid = {arXiv:2604.07297},
        pages = {arXiv:2604.07297},
          doi = {10.48550/arXiv.2604.07297},
archivePrefix = {arXiv},
       eprint = {2604.07297},
 primaryClass = {astro-ph.GA},
       adsurl = {https://ui.adsabs.harvard.edu/abs/2026arXiv260407297B}
}

@ARTICLE{thatersubm,
       author = {{Thater}, Sabine and {Chaturvedi}, Avinash and {Krajnovi{\'c}}, Davor and {Cappellari}, Michele  and {Khochfar}, Sadegh and {Naab}, Thorsten and {Sarzi}, Marc and {van de Ven}, Glenn},
        title = "{Supermassive black holes in six triaxial galaxies: Insights from SINFONI and MUSE observations}",
      journal = {\aap\ accepted},
         year = 2026,
                 month = may,
          eid={arXiv2605.28959},
           pages={arXiv2605.28959},
      archivePrefix={arXiv},
      eprint = {2605.28959},
      primaryClass={astro-ph.GA},
      url={https://arxiv.org/abs/2605.28959}, 
 }

@ARTICLE{2018MNRAS.473.3000Z,
       author = {{Zhu}, Ling and {van den Bosch}, Remco and {van de Ven}, Glenn and {Lyubenova}, Mariya and {Falc{\'o}n-Barroso}, Jes{\'u}s and {Meidt}, Sharon E. and {Martig}, Marie and {Shen}, Juntai and {Li}, Zhao-Yu and {Yildirim}, Akin and {Walcher}, C. Jakob and {Sanchez}, Sebastian F.},
        title = "{Orbital decomposition of CALIFA spiral galaxies}",
      journal = {\mnras},
         year = 2018,
        month = jan,
       volume = {473},
       number = {3},
        pages = {3000-3018},
          doi = {10.1093/mnras/stx2409},
archivePrefix = {arXiv},
       eprint = {1709.06649},
 primaryClass = {astro-ph.GA},
       adsurl = {https://ui.adsabs.harvard.edu/abs/2018MNRAS.473.3000Z}
}

@ARTICLE{2018MNRAS.479..945Z,
       author = {{Zhu}, Ling and {van de Ven}, Glenn and {M{\'e}ndez-Abreu}, Jairo and {Obreja}, Aura},
        title = "{Morphology and kinematics of orbital components in CALIFA galaxies across the Hubble sequence}",
      journal = {\mnras},
         year = 2018,
        month = sep,
       volume = {479},
       number = {1},
        pages = {945-960},
          doi = {10.1093/mnras/sty1521},
archivePrefix = {arXiv},
       eprint = {1806.02886},
 primaryClass = {astro-ph.GA},
       adsurl = {https://ui.adsabs.harvard.edu/abs/2018MNRAS.479..945Z}
}

@ARTICLE{2020MNRAS.491.1690J,
       author = {{Jin}, Yunpeng and {Zhu}, Ling and {Long}, R.~J. and {Mao}, Shude and {Wang}, Lan and {van de Ven}, Glenn},
        title = "{SDSS-IV MaNGA: Internal mass distributions and orbital structures of early-type galaxies and their dependence on environment}",
      journal = {\mnras},
         year = 2020,
        month = jan,
       volume = {491},
       number = {2},
        pages = {1690-1708},
          doi = {10.1093/mnras/stz3072},
archivePrefix = {arXiv},
       eprint = {1911.00777},
 primaryClass = {astro-ph.GA},
       adsurl = {https://ui.adsabs.harvard.edu/abs/2020MNRAS.491.1690J}
}

@ARTICLE{2019Sci...365..664D,
       author = {{Do}, Tuan and {Hees}, Aurelien and {Ghez}, Andrea and {Martinez}, Gregory D. and {Chu}, Devin S. and {Jia}, Siyao and {Sakai}, Shoko and {Lu}, Jessica R. and {Gautam}, Abhimat K. and {O'Neil}, Kelly Kosmo and {Becklin}, Eric E. and {Morris}, Mark R. and {Matthews}, Keith and {Nishiyama}, Shogo and {Campbell}, Randy and {Chappell}, Samantha and {Chen}, Zhuo and {Ciurlo}, Anna and {Dehghanfar}, Arezu and {Gallego-Cano}, Eulalia and {Kerzendorf}, Wolfgang E. and {Lyke}, James E. and {Naoz}, Smadar and {Saida}, Hiromi and {Sch{\"o}del}, Rainer and {Takahashi}, Masaaki and {Takamori}, Yohsuke and {Witzel}, Gunther and {Wizinowich}, Peter},
        title = "{Relativistic redshift of the star S0-2 orbiting the Galactic Center supermassive black hole}",
      journal = {Science},
         year = 2019,
        month = aug,
       volume = {365},
       number = {6454},
        pages = {664-668},
          doi = {10.1126/science.aav8137},
archivePrefix = {arXiv},
       eprint = {1907.10731},
 primaryClass = {astro-ph.GA},
       adsurl = {https://ui.adsabs.harvard.edu/abs/2019Sci...365..664D}
}

@ARTICLE{2019MNRAS.486.4753J,
       author = {{Jin}, Yunpeng and {Zhu}, Ling and {Long}, R.~J. and {Mao}, Shude and {Xu}, Dandan and {Li}, Hongyu and {van de Ven}, Glenn},
        title = "{Evaluating the ability of triaxial Schwarzschild modelling to estimate properties of galaxies from the Illustris simulation}",
      journal = {\mnras},
         year = 2019,
        month = jul,
       volume = {486},
       number = {4},
        pages = {4753-4772},
          doi = {10.1093/mnras/stz1170},
archivePrefix = {arXiv},
       eprint = {1904.12942},
 primaryClass = {astro-ph.GA},
       adsurl = {https://ui.adsabs.harvard.edu/abs/2019MNRAS.486.4753J}
}

@ARTICLE{2025A&A...700A.249J,
       author = {{Jin}, Yunpeng and {Zhu}, Ling and {Tahmasebzadeh}, Behzad and {Mao}, Shude and {van de Ven}, Glenn and {Guo}, Rui and {Cai}, Runsheng},
        title = "{Recovering the pattern speeds of edge-on barred galaxies via an orbit-superposition method}",
      journal = {\aap},
         year = 2025,
        month = aug,
       volume = {700},
          eid = {A249},
        pages = {A249},
          doi = {10.1051/0004-6361/202555378},
archivePrefix = {arXiv},
       eprint = {2505.02917},
 primaryClass = {astro-ph.GA},
       adsurl = {https://ui.adsabs.harvard.edu/abs/2025A&A...700A.249J}
}

@ARTICLE{2015MNRAS.447..948C,
       author = {{Chatzopoulos}, S. and {Fritz}, T.~K. and {Gerhard}, O. and {Gillessen}, S. and {Wegg}, C. and {Genzel}, R. and {Pfuhl}, O.},
        title = "{The old nuclear star cluster in the Milky Way: dynamics, mass, statistical parallax, and black hole mass}",
      journal = {\mnras},
         year = 2015,
        month = feb,
       volume = {447},
       number = {1},
        pages = {948-968},
          doi = {10.1093/mnras/stu2452},
archivePrefix = {arXiv},
       eprint = {1403.5266},
 primaryClass = {astro-ph.GA},
       adsurl = {https://ui.adsabs.harvard.edu/abs/2015MNRAS.447..948C}
}

@ARTICLE{2014MNRAS.441.3359D,
       author = {{Dutton}, Aaron A. and {Macci{\`o}}, Andrea V.},
        title = "{Cold dark matter haloes in the Planck era: evolution of structural parameters for Einasto and NFW profiles}",
      journal = {\mnras},
         year = 2014,
        month = jul,
       volume = {441},
       number = {4},
        pages = {3359-3374},
          doi = {10.1093/mnras/stu742},
archivePrefix = {arXiv},
       eprint = {1402.7073},
 primaryClass = {astro-ph.CO},
       adsurl = {https://ui.adsabs.harvard.edu/abs/2014MNRAS.441.3359D}
}

@ARTICLE{2008ApJ...682..841C,
       author = {{Chanam{\'e}}, Julio and {Kleyna}, Jan and {van der Marel}, Roeland},
        title = "{Constraining the Mass Profiles of Stellar Systems: Schwarzschild Modeling of Discrete Velocity Data Sets}",
      journal = {\apj},
         year = 2008,
        month = aug,
       volume = {682},
       number = {2},
        pages = {841-860},
          doi = {10.1086/589429},
archivePrefix = {arXiv},
       eprint = {0710.1872},
 primaryClass = {astro-ph},
       adsurl = {https://ui.adsabs.harvard.edu/abs/2008ApJ...682..841C}
}

@ARTICLE{2008MNRAS.385..647V,
       author = {{van den Bosch}, R.~C.~E. and {van de Ven}, G. and {Verolme}, E.~K. and {Cappellari}, M. and {de Zeeuw}, P.~T.},
        title = "{Triaxial orbit based galaxy models with an application to the (apparent) decoupled core galaxy NGC 4365}",
      journal = {\mnras},
         year = 2008,
        month = apr,
       volume = {385},
       number = {2},
        pages = {647-666},
          doi = {10.1111/j.1365-2966.2008.12874.x},
archivePrefix = {arXiv},
       eprint = {0712.0113},
 primaryClass = {astro-ph},
       adsurl = {https://ui.adsabs.harvard.edu/abs/2008MNRAS.385..647V}
}

@ARTICLE{1996ApJ...462..563N,
       author = {{Navarro}, Julio F. and {Frenk}, Carlos S. and {White}, Simon D.~M.},
        title = "{The Structure of Cold Dark Matter Halos}",
      journal = {\apj},
         year = 1996,
        month = may,
       volume = {462},
        pages = {563},
          doi = {10.1086/177173},
archivePrefix = {arXiv},
       eprint = {astro-ph/9508025},
 primaryClass = {astro-ph},
       adsurl = {https://ui.adsabs.harvard.edu/abs/1996ApJ...462..563N}
}

@ARTICLE{1994A&A...285..723E,
       author = {{Emsellem}, E. and {Monnet}, G. and {Bacon}, R.},
        title = "{The multi-gaussian expansion method: a tool for building realistic photometric and kinematical models of stellar systems I. The formalism}",
      journal = {\aap},
         year = 1994,
        month = may,
       volume = {285},
        pages = {723-738},
       adsurl = {https://ui.adsabs.harvard.edu/abs/1994A&A...285..723E}
}

@ARTICLE{1979ApJ...232..236S,
       author = {{Schwarzschild}, M.},
        title = "{A numerical model for a triaxial stellar system in dynamical equilibrium.}",
      journal = {\apj},
         year = 1979,
        month = aug,
       volume = {232},
        pages = {236-247},
          doi = {10.1086/157282},
       adsurl = {https://ui.adsabs.harvard.edu/abs/1979ApJ...232..236S}
}

@ARTICLE{1922MNRAS..82..122J,
       author = {{Jeans}, J.~H.},
        title = "{The Motions of Stars in a Kapteyn Universe}",
      journal = {\mnras},
         year = 1922,
        month = jan,
       volume = {82},
        pages = {122-132},
          doi = {10.1093/mnras/82.3.122},
       adsurl = {https://ui.adsabs.harvard.edu/abs/1922MNRAS..82..122J}
}

@ARTICLE{2025A&A...699A.239F,
       author = {{Feldmeier-Krause}, A. and {Ver{\v{s}}i{\v{c}}}, T. and {van de Ven}, G. and {Gallego-Cano}, E. and {Neumayer}, N.},
        title = "{Dynamical mass distribution and velocity structure of the Galactic centre}",
      journal = {\aap},
         year = 2025,
        month = jul,
       volume = {699},
          eid = {A239},
        pages = {A239},
          doi = {10.1051/0004-6361/202554630},
archivePrefix = {arXiv},
       eprint = {2506.06014},
 primaryClass = {astro-ph.GA},
       adsurl = {https://ui.adsabs.harvard.edu/abs/2025A&A...699A.239F}
}

@ARTICLE{2002MNRAS.333..400C,
       author = {{Cappellari}, Michele},
        title = "{Efficient multi-Gaussian expansion of galaxies}",
      journal = {\mnras},
         year = 2002,
        month = jun,
       volume = {333},
       number = {2},
        pages = {400-410},
          doi = {10.1046/j.1365-8711.2002.05412.x},
archivePrefix = {arXiv},
       eprint = {astro-ph/0201430},
 primaryClass = {astro-ph},
       adsurl = {https://ui.adsabs.harvard.edu/abs/2002MNRAS.333..400C}
}

@ARTICLE{2019MNRAS.482.1417B,
       author = {{Bennett}, Morgan and {Bovy}, Jo},
        title = "{Vertical waves in the solar neighbourhood in Gaia DR2}",
      journal = {\mnras},
         year = 2019,
        month = jan,
       volume = {482},
       number = {1},
        pages = {1417-1425},
          doi = {10.1093/mnras/sty2813},
archivePrefix = {arXiv},
       eprint = {1809.03507},
 primaryClass = {astro-ph.GA},
       adsurl = {https://ui.adsabs.harvard.edu/abs/2019MNRAS.482.1417B}
}

@ARTICLE{2023A&A...675A..18T,
       author = {{Thater}, Sabine and {Lyubenova}, Mariya and {Fahrion}, Katja and {Mart{\'\i}n-Navarro}, Ignacio and {Jethwa}, Prashin and {Nguyen}, Dieu D. and {van de Ven}, Glenn},
        title = "{Effect of the initial mass function on the dynamical SMBH mass estimate in the nucleated early-type galaxy FCC 47}",
      journal = {\aap},
         year = 2023,
        month = jul,
       volume = {675},
          eid = {A18},
        pages = {A18},
          doi = {10.1051/0004-6361/202245362},
archivePrefix = {arXiv},
       eprint = {2304.13310},
 primaryClass = {astro-ph.GA},
       adsurl = {https://ui.adsabs.harvard.edu/abs/2023A&A...675A..18T}
}

@ARTICLE{2022A&A...667A..51T,
       author = {{Thater}, Sabine and {Jethwa}, Prashin and {Tahmasebzadeh}, Behzad and {Zhu}, Ling and {den Brok}, Mark and {Santucci}, Giulia and {Ding}, Yuchen and {Poci}, Adriano and {Lilley}, Edward and {Tim de Zeeuw}, P. and {Zocchi}, Alice and {Maindl}, Thomas I. and {Rigamonti}, Fabio and {Yang}, Meng and {Fahrion}, Katja and {van de Ven}, Glenn},
        title = "{Testing the robustness of DYNAMITE triaxial Schwarzschild modelling: The effects of correcting the orbit mirroring}",
      journal = {\aap},
         year = 2022,
        month = nov,
       volume = {667},
          eid = {A51},
        pages = {A51},
          doi = {10.1051/0004-6361/202243926},
archivePrefix = {arXiv},
       eprint = {2205.04165},
 primaryClass = {astro-ph.GA},
       adsurl = {https://ui.adsabs.harvard.edu/abs/2022A&A...667A..51T}
}

@ARTICLE{2013MNRAS.431.3364L,
       author = {{Lyubenova}, Mariya and {van den Bosch}, Remco C.~E. and {C{\^o}t{\'e}}, Patrick and {Kuntschner}, Harald and {van de Ven}, Glenn and {Ferrarese}, Laura and {Jord{\'a}n}, Andr{\'e}s and {Infante}, Leopoldo and {Peng}, Eric W.},
        title = "{The complex nature of the nuclear star cluster in FCC 277}",
      journal = {\mnras},
         year = 2013,
        month = jun,
       volume = {431},
       number = {4},
        pages = {3364-3372},
          doi = {10.1093/mnras/stt414},
archivePrefix = {arXiv},
       eprint = {1303.1210},
 primaryClass = {astro-ph.CO},
       adsurl = {https://ui.adsabs.harvard.edu/abs/2013MNRAS.431.3364L}
}

@ARTICLE{2013MNRAS.434.3174V,
       author = {{Vasiliev}, Eugene},
        title = "{A new code for orbit analysis and Schwarzschild modelling of triaxial stellar systems}",
      journal = {\mnras},
         year = 2013,
        month = oct,
       volume = {434},
       number = {4},
        pages = {3174-3195},
          doi = {10.1093/mnras/stt1235},
archivePrefix = {arXiv},
       eprint = {1307.8116},
 primaryClass = {astro-ph.GA},
       adsurl = {https://ui.adsabs.harvard.edu/abs/2013MNRAS.434.3174V}
}

@ARTICLE{2020ApJ...889...39V,
       author = {{Vasiliev}, Eugene and {Valluri}, Monica},
        title = "{A New Implementation of the Schwarzchild Method for Constructing Observationally Driven Dynamical Models of Galaxies of All Morphological Types}",
      journal = {\apj},
         year = 2020,
        month = jan,
       volume = {889},
       number = {1},
          eid = {39},
        pages = {39},
          doi = {10.3847/1538-4357/ab5fe0},
archivePrefix = {arXiv},
       eprint = {1912.04288},
 primaryClass = {astro-ph.GA},
       adsurl = {https://ui.adsabs.harvard.edu/abs/2020ApJ...889...39V}
}

@ARTICLE{2000ApJ...539L..13G,
       author = {{Gebhardt}, Karl and {Bender}, Ralf and {Bower}, Gary and {Dressler}, Alan and {Faber}, S.~M. and {Filippenko}, Alexei V. and {Green}, Richard and {Grillmair}, Carl and {Ho}, Luis C. and {Kormendy}, John and {Lauer}, Tod R. and {Magorrian}, John and {Pinkney}, Jason and {Richstone}, Douglas and {Tremaine}, Scott},
        title = "{A Relationship between Nuclear Black Hole Mass and Galaxy Velocity Dispersion}",
      journal = {\apjl},
         year = 2000,
        month = aug,
       volume = {539},
       number = {1},
        pages = {L13-L16},
          doi = {10.1086/312840},
archivePrefix = {arXiv},
       eprint = {astro-ph/0006289},
 primaryClass = {astro-ph},
       adsurl = {https://ui.adsabs.harvard.edu/abs/2000ApJ...539L..13G}
}

@ARTICLE{2000ApJ...539L...9F,
       author = {{Ferrarese}, Laura and {Merritt}, David},
        title = "{A Fundamental Relation between Supermassive Black Holes and Their Host Galaxies}",
      journal = {\apjl},
         year = 2000,
        month = aug,
       volume = {539},
       number = {1},
        pages = {L9-L12},
          doi = {10.1086/312838},
archivePrefix = {arXiv},
       eprint = {astro-ph/0006053},
 primaryClass = {astro-ph},
       adsurl = {https://ui.adsabs.harvard.edu/abs/2000ApJ...539L...9F}
}

@ARTICLE{2013ApJ...763...76S,
       author = {{Scott}, Nicholas and {Graham}, Alister W.},
        title = "{Updated Mass Scaling Relations for Nuclear Star Clusters and a Comparison to Supermassive Black Holes}",
      journal = {\apj},
         year = 2013,
        month = feb,
       volume = {763},
       number = {2},
          eid = {76},
        pages = {76},
          doi = {10.1088/0004-637X/763/2/76},
archivePrefix = {arXiv},
       eprint = {1205.5338},
 primaryClass = {astro-ph.CO},
       adsurl = {https://ui.adsabs.harvard.edu/abs/2013ApJ...763...76S}
}

@ARTICLE{2006ApJ...644L..21F,
       author = {{Ferrarese}, Laura and {C{\^o}t{\'e}}, Patrick and {Dalla Bont{\`a}}, Elena and {Peng}, Eric W. and {Merritt}, David and {Jord{\'a}n}, Andr{\'e}s and {Blakeslee}, John P. and {Ha{\c{s}}egan}, Monica and {Mei}, Simona and {Piatek}, Slawomir and {Tonry}, John L. and {West}, Michael J.},
        title = "{A Fundamental Relation between Compact Stellar Nuclei, Supermassive Black Holes, and Their Host Galaxies}",
      journal = {\apjl},
         year = 2006,
        month = jun,
       volume = {644},
       number = {1},
        pages = {L21-L24},
          doi = {10.1086/505388},
archivePrefix = {arXiv},
       eprint = {astro-ph/0603840},
 primaryClass = {astro-ph},
       adsurl = {https://ui.adsabs.harvard.edu/abs/2006ApJ...644L..21F}
}

@ARTICLE{2009ApJ...690.1031B,
       author = {{Barth}, Aaron J. and {Strigari}, Louis E. and {Bentz}, Misty C. and {Greene}, Jenny E. and {Ho}, Luis C.},
        title = "{Dynamical Constraints on the Masses of the Nuclear Star Cluster and Black Hole in the Late-Type Spiral Galaxy NGC 3621}",
      journal = {\apj},
         year = 2009,
        month = jan,
       volume = {690},
       number = {1},
        pages = {1031-1044},
          doi = {10.1088/0004-637X/690/1/1031},
archivePrefix = {arXiv},
       eprint = {0809.1066},
 primaryClass = {astro-ph},
       adsurl = {https://ui.adsabs.harvard.edu/abs/2009ApJ...690.1031B}
}

@ARTICLE{2018ApJ...858..118N,
       author = {{Nguyen}, Dieu D. and {Seth}, Anil C. and {Neumayer}, Nadine and {Kamann}, Sebastian and {Voggel}, Karina T. and {Cappellari}, Michele and {Picotti}, Arianna and {Nguyen}, Phuong M. and {B{\"o}ker}, Torsten and {Debattista}, Victor and {Caldwell}, Nelson and {McDermid}, Richard and {Bastian}, Nathan and {Ahn}, Christopher C. and {Pechetti}, Renuka},
        title = "{Nearby Early-type Galactic Nuclei at High Resolution: Dynamical Black Hole and Nuclear Star Cluster Mass Measurements}",
      journal = {\apj},
         year = 2018,
        month = may,
       volume = {858},
       number = {2},
          eid = {118},
        pages = {118},
          doi = {10.3847/1538-4357/aabe28},
archivePrefix = {arXiv},
       eprint = {1711.04314},
 primaryClass = {astro-ph.GA},
       adsurl = {https://ui.adsabs.harvard.edu/abs/2018ApJ...858..118N}
}

@ARTICLE{2022A&A...660A.133H,
       author = {{Hermosa Mu{\~n}oz}, L. and {M{\'a}rquez}, I. and {Cazzoli}, S. and {Masegosa}, J. and {Ag{\'\i}s-Gonz{\'a}lez}, B.},
        title = "{A search for ionised gas outflows in an H{\ensuremath{\alpha}} imaging atlas of nearby LINERs}",
      journal = {\aap},
         year = 2022,
        month = apr,
       volume = {660},
          eid = {A133},
        pages = {A133},
          doi = {10.1051/0004-6361/202142629},
archivePrefix = {arXiv},
       eprint = {2201.05080},
 primaryClass = {astro-ph.GA},
       adsurl = {https://ui.adsabs.harvard.edu/abs/2022A&A...660A.133H}
}

@ARTICLE{2022MNRAS.509.5416T,
       author = {{Thater}, Sabine and {Krajnovi{\'c}}, Davor and {Weilbacher}, Peter M. and {Nguyen}, Dieu D. and {Bureau}, Martin and {Cappellari}, Michele and {Davis}, Timothy A. and {Iguchi}, Satoru and {McDermid}, Richard and {Onishi}, Kyoko and {Sarzi}, Marc and {van de Ven}, Glenn},
        title = "{Cross-checking SMBH mass estimates in NGC 6958 - I. Stellar dynamics from adaptive optics-assisted MUSE observations}",
      journal = {\mnras},
         year = 2022,
        month = feb,
       volume = {509},
       number = {4},
        pages = {5416-5436},
          doi = {10.1093/mnras/stab3210},
archivePrefix = {arXiv},
       eprint = {2111.01620},
 primaryClass = {astro-ph.GA},
       adsurl = {https://ui.adsabs.harvard.edu/abs/2022MNRAS.509.5416T}
}

@ARTICLE{2012ApJ...753...79W,
       author = {{Walsh}, Jonelle L. and {van den Bosch}, Remco C.~E. and {Barth}, Aaron J. and {Sarzi}, Marc},
        title = "{A Stellar Dynamical Mass Measurement of the Black Hole in NGC 3998 from Keck Adaptive Optics Observations}",
      journal = {\apj},
         year = 2012,
        month = jul,
       volume = {753},
       number = {1},
          eid = {79},
        pages = {79},
          doi = {10.1088/0004-637X/753/1/79},
archivePrefix = {arXiv},
       eprint = {1205.0816},
 primaryClass = {astro-ph.CO},
       adsurl = {https://ui.adsabs.harvard.edu/abs/2012ApJ...753...79W}
}

@Article{Hunter:2007, Author = {Hunter, J. D.}, Title = {Matplotlib: A 2D graphics environment}, Journal = {Computing In Science \& Engineering}, Volume = {9}, Number = {3}, Pages = {90--95}, publisher = {IEEE COMPUTER SOC}, year = 2007 }

\begin{appendix} 
  \section{Corner plots}
\begin{figure*}
 \centering
 \includegraphics[width=18cm]{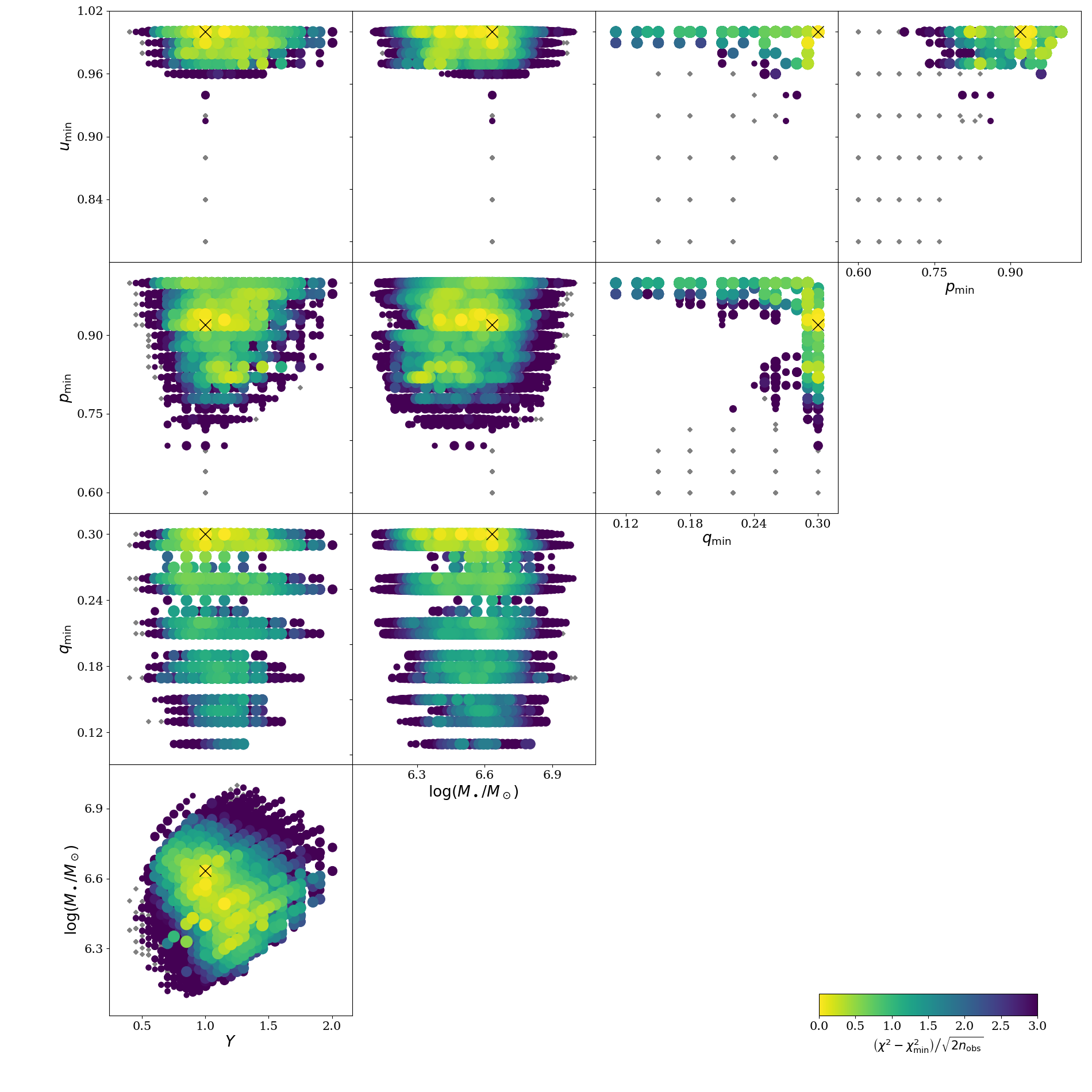}
      \caption{Illustration of the sampled parameter space, each point denotes one of \textgreater 14\,000 models. The symbol size and colour represent the value of $\chi^2$, as indicated by the colour bar; the black x-symbol denotes the best-fitting model. The parameters are, from top to bottom: \umin, \pmin, \qmin, \mbh, and \ups.
              }
         \label{fig:kinmapchi2}
   \end{figure*}

\begin{figure*}
 \centering
 \includegraphics[width=18cm]{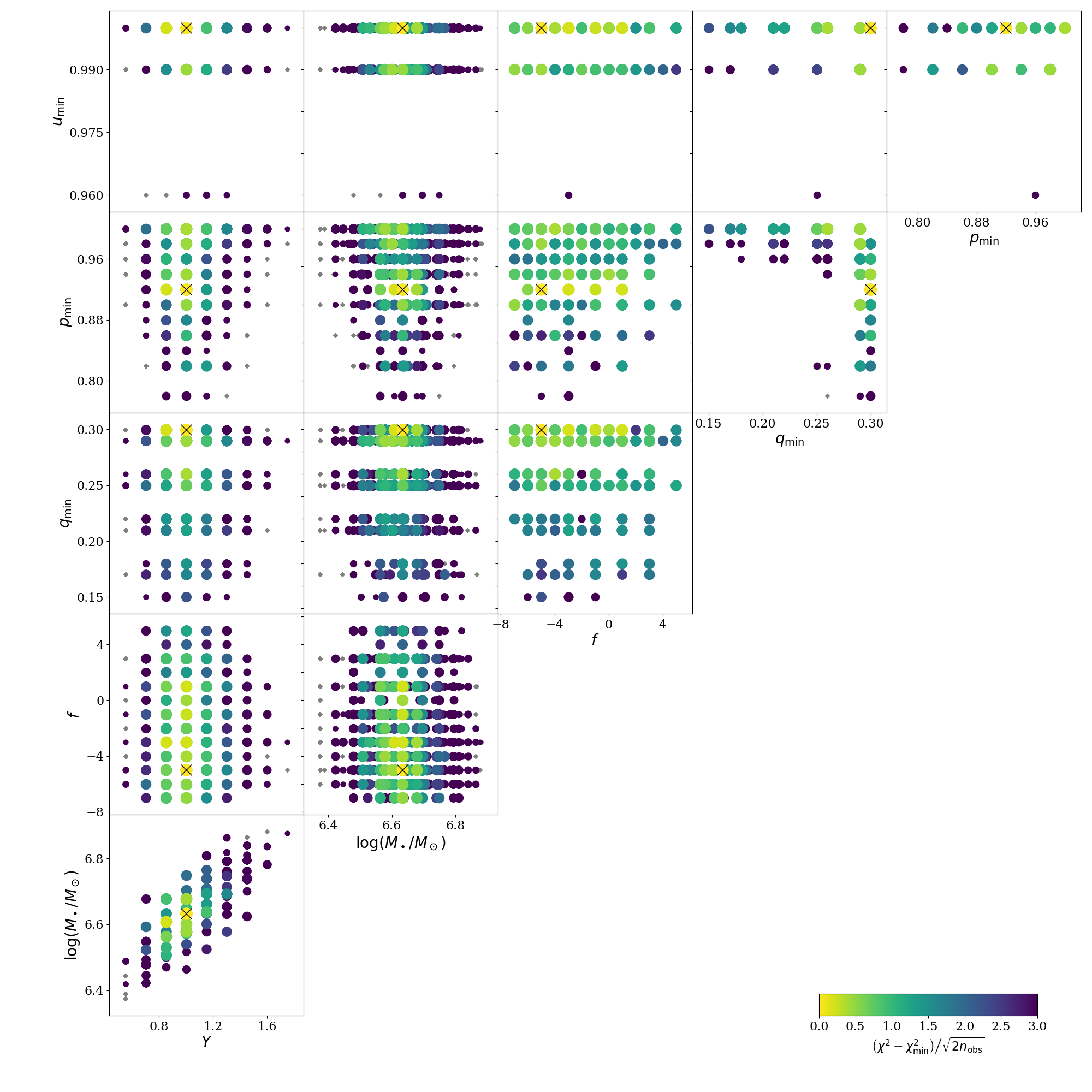}
      \caption{Same as Fig.~\ref{fig:kinmapchi2}, but for models with a spherical DM component and a more narrow range of \mbh. The parameters are, from top to bottom: \umin, \pmin, \qmin, $f$, \mbh, and \ups.
              }
         \label{fig:kinmapchi2_dm}
   \end{figure*}

  We show the parameter space spanned by our models and the location of the best-fitting model in Fig.~\ref{fig:kinmapchi2} (14\,000 models without DM) and in Fig.~\ref{fig:kinmapchi2_dm} (3\,000 models with DM). The shape parameter \qmin is near the edge of the grid, and \umin is close to 1, indicating that near edge-on models are preferred, with the major axis being parallel to the Galactic plane.

 \section{$R_\text{max}$ versus $z_\text{max}$ plots}
 \label{sec:rmaxzmax}
\begin{figure}
 \includegraphics[width=0.87\columnwidth]{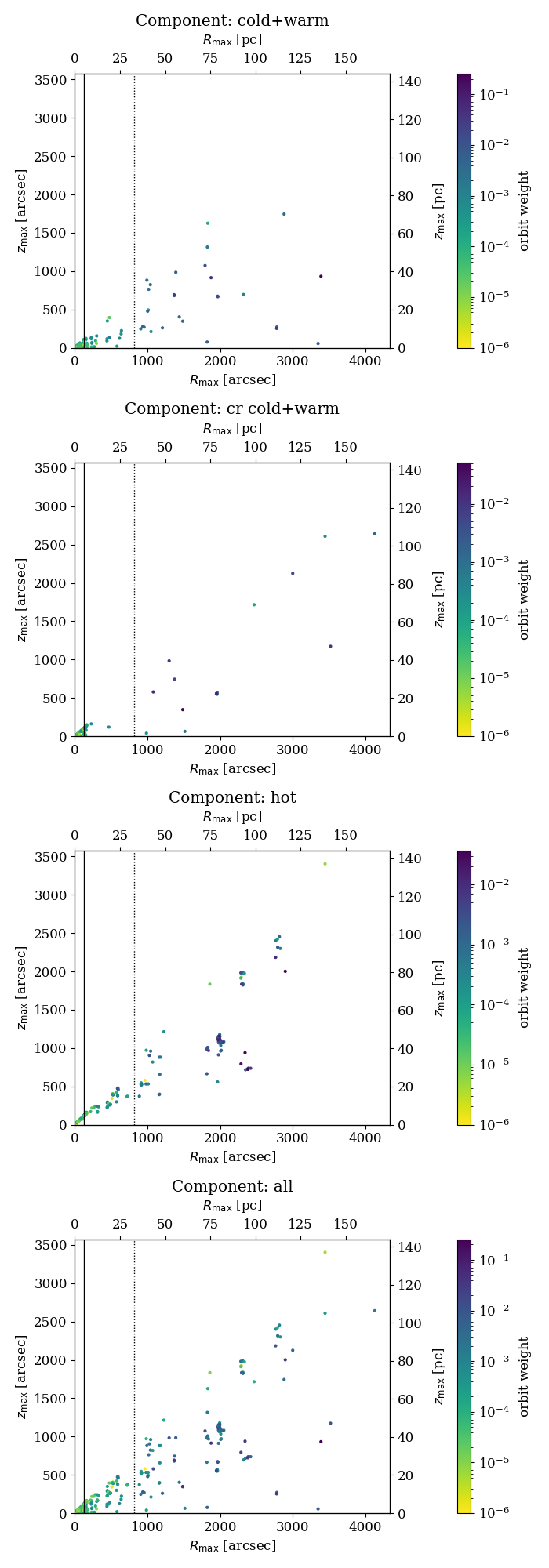}

  \caption{$R_\text{max}$ vs. $z_\text{max}$ diagram of the best-fit model. Each data point represents an orbit with non-zero weight, the colour represents the relative weight of the orbit, the solid line denotes 1~$R_e$ of the NSC, and the dashed line denotes the outer extent of the kinematic data. The panels show, from top to bottom, regular cold and warm tube orbits ($\lambda_z$\textgreater 0.25), CR orbits ($\lambda_z\leq$-0.25), hot orbits, and all orbits. }
  \label{fig:rmaxzmaxout}
\end{figure}

\begin{figure}
 \includegraphics[width=0.87\columnwidth]{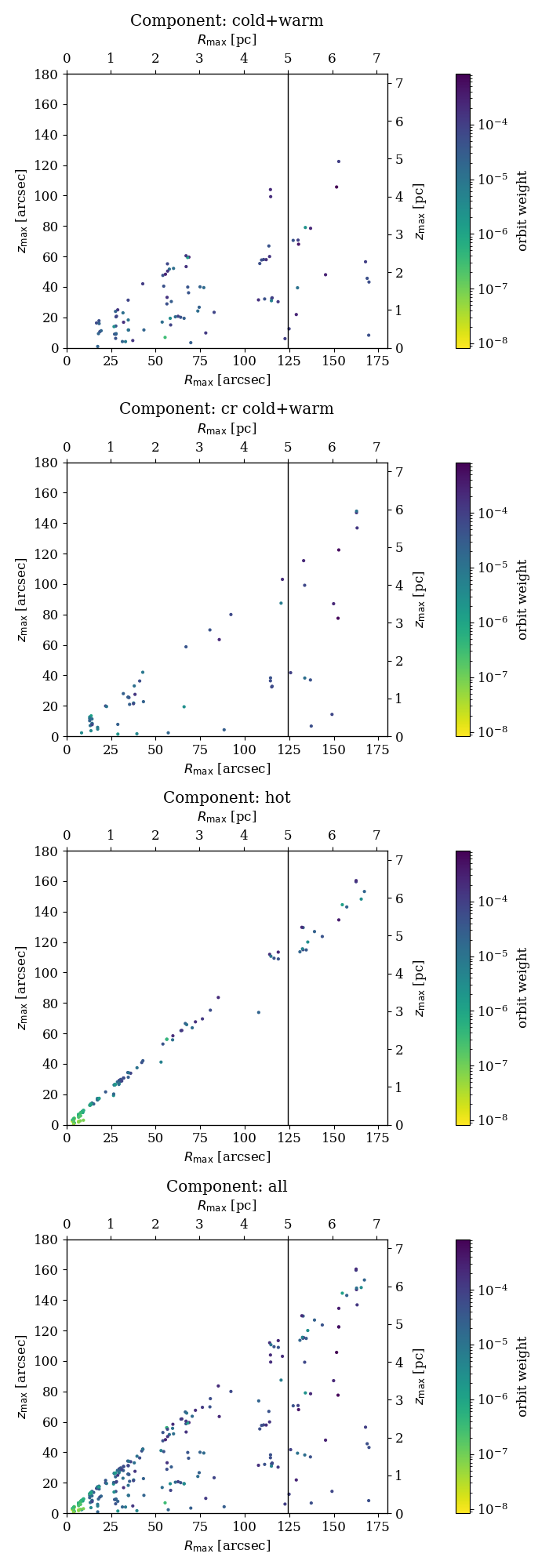}

  \caption{Same as Fig.~\ref{fig:rmaxzmaxout}, but zoomed into the region dominated by the NSC. Note the different colour scale to improve visibility of a range of orbit weights.}
  \label{fig:rmaxzmaxin}
\end{figure}

 In Figs.~\ref{fig:rmaxzmaxout} and \ref{fig:rmaxzmaxin}, we show the distribution of orbit weights of the best-fit model in $R_\text{max}$ vs $z_\text{max}$ space. $R_\text{max}$ denotes the maximum of the orbits' intrinsic radii, while $z_\text{max}$ denotes the orbits' maximum distance to the major axis. Hot orbits have higher values of $z_\text{max}$ than cold orbits. The orbits in the inner region are finer sampled, as indicated by the higher density of points in the inner $\lesssim$33\,pc region.


 \section{Velocity anisotropy}

\begin{figure}
\includegraphics[width=\columnwidth]{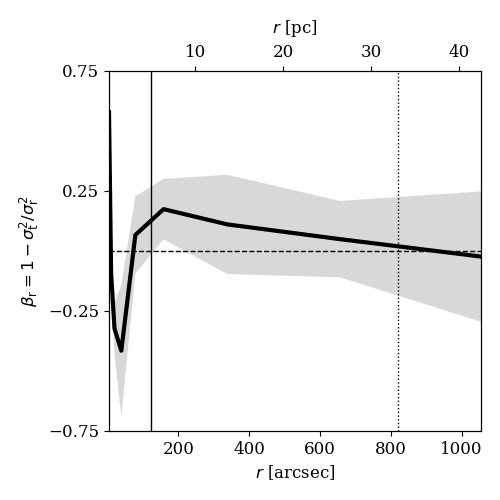}
  \caption{Radial velocity anisotropy $\beta_r$ as function of intrinsic radius. The black line denotes the best-fit model without DM, the shaded region the 1$\sigma$ uncertainty, the vertical solid line 1\,\re\ of the NSC, the vertical dotted line the outer limit of the kinematic data. }
  \label{fig:aniso}
\end{figure}

The intrinsic velocity anisotropy profile $\beta_r$ is shown in Fig.~\ref{fig:aniso}. 
Anisotropy is defined as $\beta_r$=1-$\sigma_t^2/\sigma_r^2$ ($\sigma_r, \sigma_\phi, \sigma_\theta$ are the radial, azimuthal angular and polar angular velocity dispersion in spherical co-ordinates\footnote{$\sigma_t=\sqrt{\sigma_\phi^2 + \sigma_\theta^2}$}), plotted along intrinsic radius, $r$. 
A value of zero denotes an isotropic velocity distribution; a higher (lower) value than zero denotes radially (tangentially) anisotropic orbits. Radial anisotropy is a sign of dynamically hot orbits, and tangential anisotropy indicates more circular orbits.  Our models indicate some tangential anisotropy within the inner $\sim$3\,pc of the NSC, but are in agreement with isotropy for \textgreater 10\,pc. 

\end{appendix}

\end{document}